\documentclass[fp,12pt]{jpsj3}

\usepackage{bm}
\usepackage{color,xcolor}
\usepackage{soul}
\usepackage{physics}
\usepackage{placeins}
\usepackage{amsmath,amssymb}
\usepackage{graphicx,multirow,tabularx}
\usepackage[ruled,vlined]{algorithm2e}
\usepackage{lineno}

\usepackage{comment}
\usepackage{amsmath,amssymb}
\usepackage{mathtools}
\usepackage{txfonts}

\makeatletter
\providecommand\ext@figure{lof}

\providecommand\ext@table{lot}
\makeatother
\usepackage[colorlinks=true, pdfstartview=FitV, linkcolor=magenta,citecolor=blue, urlcolor=magenta, bookmarks=true, bookmarksnumbered=true, breaklinks]{hyperref}

\newcommand{\Tc}[0]{T_\textrm{c}}
\newcommand{\nnsum}[0]{\sum_{\langle i,j \rangle}}

\title{Lee-Yang-zero ratio method in three-dimensional Ising model}

\author{
Tatsuya Wada$^1$,
Masakiyo Kitazawa$^{1,2}$, 
Kazuyuki Kanaya$^3$
}
\inst{
  $^1$Yukawa Institute for Theoretical Physics, Kyoto University, Kyoto, 606-8502, Japan\\
  $^2$J-PARC Branch, KEK Theory Center, Institute of Particle and Nuclear Studies, KEK, Tokai, Ibaraki 319-1106, Japan\\
  $^3$Tomonaga Center for the History of the Universe, University of Tsukuba, Tsukuba, Ibaraki 305-8571, Japan}

\abst{
By performing Monte Carlo simulations of the three-dimensional Ising model, we apply the recently proposed Lee-Yang-zero ratio (LYZR) method to determine the location of the critical point in this model. We demonstrate that the LYZR method is as powerful as the conventional Binder-cumulant method in studying the critical point,
while the LYZR method suppresses
the correction to finite-size scaling and non-linearity near the critical point compared to the Binder-cumulant method. We also achieve a precise determination of the values of the LYZRs at the critical point, which are universal numbers. In addition, we propose an alternative method that uses only a single Lee-Yang zero and show that it is also useful for the search for the critical point.
}

\begin{document}
\vspace*{-1.1cm}
\begin{flushright}
    \small
    YITP-25-128, J-PARC-TH-0324, UTHEP-810
\end{flushright}
\vspace*{-0.5cm}
\maketitle

\section{Introduction}
\label{sec:Introduction}

Critical phenomena are fundamental concepts across various fields of physics, including statistical mechanics, condensed-matter and high-energy physics, and cosmology. Although the physical scales involved in these fields differ by more than ten orders of magnitude, when a critical point (CP) exists, distinct physical systems exhibit identical critical behavior governed by the scaling laws and universality classes, in which symmetry and dimensionality of the system play crucial roles~\cite{WILSON197475}.
On the other hand, the existence and precise location of a critical point are not constrained by symmetries. Instead, the microscopic properties of the system, such as the system's dynamics and interactions, play essential roles.  

Various theoretical approaches have been developed to investigate critical phenomena, including the mean-field approximation~\cite{weiss:jpa-00241247,Landau:1937obd} and the renormalization-group methods~\cite{PhysicsPhysiqueFizika.2.263,PhysRevLett.28.240,PhysRevB.4.3174}. 
However, they give exact solutions only in limited systems. Therefore, in most cases, numerical methods, such as the Monte Carlo~\cite{Ferrenberg:2018zst,10.1063/1.1699114,Binder:2001ha}, tensor network~\cite{PhysRevLett.69.2863,Levin:2006jai}, and conformal-bootstrap~\cite{Rattazzi:2008pe,Chang:2024whx} methods, serve as primary tools for locating the CP and determining its properties.

In numerical analyses of critical phenomena, it is crucial to properly account for finite-size effects, as simulations are always performed on finite volumes while the correlation length diverges at the CP.
When the system size is sufficiently large, it is known that thermodynamic quantities in the vicinity of the CP obey an extended scaling law known as the finite-size scaling (FSS), where the system size acts as a scaling variable~\cite{PhysRevLett.28.1516,Binder:1981sa,Binder:2001ha,Pelissetto:2000ek}. 
The analysis based on the FSS is crucial for revealing properties of the CP, such as its location and the critical exponents.
A useful method among them is the so-called Binder-cumulant method~\cite{Binder:1981sa},
in which the location of the CP is determined from the intersection point of the ratios of cumulants obtained at various system sizes.

In our recent study~\cite{Wada:2024qsk}, we proposed a new systematic way to locate the CP based on the FSS. This method makes use of the Lee-Yang zeros (LYZs)~\cite{Lee:1952ig,Yang:1952be}, i.e.\ zeros of the partition function on the complex-parameter space. 
Characteristic behavior of LYZs near the CP has been investigated in the literature~\cite{Itzykson:1983gb,Butera:2012tq,Gliozzi:2013ysa,he2025revisitingfermionsignproblem,liu2025determinationmeltingtemperaturehexagonal,Wan:2025wdg,shastry2025partition,abdelshafy2025yang,song2025new,honchar2025partition,Rennecke:2022ohx,Connelly:2020gwa,Singh:2023bog,Karsch:2023rfb}. In particular, the edge of the distribution of the LYZ, called the Lee-Yang edge singularity~\cite{Fisher:1978pf}, is known to follow a specific scaling law. This scaling property has been recently applied for the numerical search for the CP in quantum chromodynamics (QCD) at non-zero chemical potential in lattice QCD Monte Carlo simulations~\cite{Dimopoulos:2021vrk,Ding:2024sux,Clarke:2024ugt,Basar:2023nkp,Schmidt:2025ppy,Adam:2025phc}. However, the finite-size effects are not taken into account in these analyses.

In order to properly control finite-size effects exploiting the FSS of the LYZs, in Ref.~\citen{Wada:2024qsk} we proposed the use of the ratios of LYZs. We showed that the ratios of the imaginary parts of LYZs possess similar properties as the Binder cumulants, i.e.\ the ratios obtained on various system sizes intersect at the CP. The LYZ ratios (LYZRs) thus serve as an alternative and independent approach to locating the CP numerically through intersection analysis. In Ref.~\citen{Wada:2024qsk}, based on the FSS we also showed that this property of LYZRs applies to any CP in general systems, and the values of LYZRs at the CP are universal numbers that are specific for individual universality classes. Compared with the Binder-cumulant method, this method, which we call the LYZR method in what follows, has characteristics that the convergence of the finite-size effects is faster in general systems. We have also verified the validity of this method through a numerical analysis of the CP in the three-dimensional three-state Potts model with an external field, which belongs to the same universality class as the three-dimensional Ising model~\cite{Potts_1952,Karsch:2000xv}. It has been shown that the LYZR method can successfully determine the location and the critical exponents of the CP in this model. 
Preliminary results of the application of the LYZR method to QCD in the heavy-quark region are also presented in Ref.~\citen{Wada:2024lat}.

In the present study, we apply the LYZR method to Monte Carlo simulations of the CP in the three-dimensional Ising model. As the critical temperature $T_{\rm c}$ has been measured with great precision~\cite{Ferrenberg:2018zst}, this model serves as a good testing ground for the LYZR method and its comparison with other methods such as the Binder-cumulant method. We demonstrate that the LYZR method can determine the value of $T_{\rm c}$ and other parameters with nearly the same precision as the Binder-cumulant method. We also show that the method suppresses the correction to FSS and nonlinearities in the temperature dependence. Precise analyses of the values of LYZRs at the CP, which are the universal numbers, are also performed utilizing the simplicity of the model. 

In addition, we propose a series of alternative methods to realize the intersection analysis of the CP with the use of a single LYZ. Whereas these methods assume that the universality class and the values of critical exponents are known, they are applicable to systems where the search for the second LYZ is difficult, such as the lattice-QCD simulations~\cite{Dimopoulos:2021vrk,Ding:2024sux,Clarke:2024ugt,Basar:2023nkp,Schmidt:2025ppy,Adam:2025phc}. 

This paper is organized as follows.  
In Sec.~\ref{sec:formalism}, we introduce the LYZR method and compare it with the Binder-cumulant method. The analysis with a single LYZ is also introduced. After describing the setup of our numerical simulations in Sec.~\ref{sec:setup}, we present our numerical results of the LYZR and Binder-cumulant methods in Sec.~\ref{sec:Numerical_result}. The numerical results for the single-LYZ method are reported in Sec.~\ref{sec:Single_result}.
In Sec.~\ref{sec:Discussion}, we give a brief summary.

\section{Theoretical framework}
\label{sec:formalism}

\subsection{Ising model}
\label{sec:model}

Throughout this study, we investigate the ferromagnetic three-dimensional Ising (3d-Ising) model~\cite{Ising:1925em}
\begin{align}
    H(h,L;\{\sigma\}) = - J \nnsum \sigma_i \sigma_j - h\sum_{i} \sigma_i ,
    \label{eq:def_Ising_H}
\end{align}
with the interaction constant $J>0$ and external magnetic field $h$ on the cubic lattice of size $L^3$ under the periodic boundary conditions in all directions. Each lattice site $i$ has spin $\sigma_i$, which takes on the values $\sigma_i=\pm1$, and $\langle i,j \rangle$ represents pairs of nearest-neighbor sites. In the following, we set $J=1$ to make all physical quantities dimensionless.

The 3d-Ising model~\eqref{eq:def_Ising_H} at nonzero temperature $T$ in the $L\to\infty$ limit exhibits a CP at $(T,h)=(\Tc,0)$, which is the endpoint of the ferromagnetic first-order phase-transition line at $h=0$ and $T<\Tc$. The most precise value of $\Tc$ known so far, to the best of the authors' knowledge, is obtained by a Monte Carlo simulation as~\cite{Ferrenberg:2018zst}
\begin{align}
    \Tc = 4.51152321(4). 
    \label{eq:Tc}
\end{align}
In the following, we use the reduced temperature 
\begin{align}
    t = \frac{T-\Tc}{\Tc}  .
    \label{eq:reduced_T}
\end{align}
Conversion between $T$ and $t$ in the arguments of the following functions is understood.

\subsection{Lee-Yang zeros}
\label{sec:LYZ}

The partition function of the 3d-Ising model is given by
\begin{align}
    Z(t,h,L) 
    = \sum_{\{\sigma_i\}} e^{-H(h,L;\{\sigma\})/T} 
    = \sum_{\{\sigma_i\}} \exp\bigg(\frac{1}{T}\nnsum \sigma_i \sigma_j + \frac{h}{T}\sum_{i}  \sigma_i\bigg)\,.
    \label{eq:def_Ising_Z}
\end{align} 
While it is evident from Eq.~\eqref{eq:def_Ising_Z} that $Z(t,h,L)$ is real and satisfies $Z(t,h,L)>0$ for real $T$ and $h$, it may vanish for complex parameters $T\in\mathbb{C}$ and/or $h\in\mathbb{C}$. Among them, the zeros on the complex-$h$ plane for real $T$ are called the LYZs~\cite{Yang:1952be,Lee:1952ig}. According to the Lee–Yang theorem, the LYZs in the 3d-Ising model are purely imaginary; that is, they always lie on the imaginary-$h$ axis.~\cite{Lee:1952ig}. 

Since the total magnetization 
\begin{align}
    M = \sum_{i} \sigma_i ,
    \label{eq:M}
\end{align}
takes only even or odd integer values in the range $-L^3\le M\le L^3$, Eq.~\eqref{eq:def_Ising_Z} is represented as a polynomial of the fugacity squared $e^{2h/T}$. The partition function can thus be factorized as $Z(t,h,L) \sim e^{-L^3h/T}\prod_n ( e^{2h/T} - c_n(t))$ with $L^3$ factors, where $|c_n(t)|=1$ by the Lee-Ynag theorem~\cite{Yang:1952be}. This means that the number of LYZs for $-\pi<2{\rm Im}\, h/T\le \pi$ is $L^3$ with possible degeneracies; hence they are discretely distributed.
In the following, we denote the $n$th closest LYZs from the real-$h$ axis for ${\rm Im}\,h>0$ as 
\begin{align}
    h = h_\textrm{LY}^{(n)}(t,L), \label{eq:Ising_LYZs}
\end{align}
which are pure imaginary functions of $t$ and $L$.
Because Eq.~\eqref{eq:def_Ising_Z} is real for $t \in \mathbb{R}$ and $h \in \mathbb{R}$, Eq.~\eqref{eq:def_Ising_Z} has LYZs also at $h = [h_\textrm{LY}^{(n)}(t,L)]^* = -h_\textrm{LY}^{(n)}(t,L)$ for ${\rm Im}\,h<0$ from the Schwarz reflection principle.

Let us inspect the analyticity of Eq.~\eqref{eq:Ising_LYZs}, as this property plays a key role in later discussions. Since $Z(t,h,L)$ is an analytic function of $t$ and $h$ for finite $L$, it is suggested that $h_\textrm{LY}^{(n)}(t,L)$ should also be analytic in $t$; otherwise $Z(t,h,L)$ will not be analytic in $t$. This argument should hold as long as the LYZs do not have degeneracies. However, the situation is subtle when LYZs are degenerated at some $t$. For example, the partition function can contain a factor of the form
$
    Z(t,h,L)=\big((h-h')^2- k(t-t')\big)\times \cdots 
$
when LYZs are degenerated at $t=t'$.
In this case, the LYZs $h=h'\pm k^{1/2}(t-t')^{1/2}$ are non-analytic at $t=t'$ even though $Z(t,h,L)$ is analytic. (To satisfy the Lee-Yang theorem, $h'$ should be pure imaginary and $k<0$. )
However, we have numerically confirmed that the LYZs near the real axis are always non-degenerate near the CP. We therefore assume that $h_\textrm{LY}^{(n)}(t,L)$ are analytic in $t$ in the following discussion.

\subsection{Finite-size scaling}
\label{sec:FSS}
In the vicinity of the CP and for sufficiently large $L$, the free energy $F(t,h,L)=-T\ln Z(t,h,L)$ can be decomposed into the singular and regular parts as $F(t,h,L)=F_{\rm sing}(t,h,L)+F_{\rm reg}(t,h,L)$, where the singular part obeys the FSS relation~\cite{PhysRevLett.28.1516,Binder:2001ha,Pelissetto:2000ek}
\begin{align}
    F_\textrm{sing}(t,h,L) = \tilde{F}(L^{y_t}t,L^{y_h}h) .
    \label{eq:FSS_F}
\end{align}
Here, $\tilde{F}$ is called the scaling function, and the critical exponents $y_t$ and $y_h$ are specific to the universality class; their values in the 3d-Ising model are known with high precision as~\cite{Chang:2024whx}
\begin{align}
     y_t = 1.58737472(29) , \qquad
     y_h = 2.481851194(24).
     \label{eq:ytyh}
\end{align}
In response to the decomposition of the free energy, the partition function near the CP can be written as $Z(t,h,L)=Z_{\rm sing}(t,h,L) \, Z_{\rm reg}(t,h,L)$, where $Z_{\rm sing}(t,h,L)$ satisfies
\begin{align}    
    Z_\textrm{sing}(t,h,L) = \tilde{Z}(L^{y_t}t,L^{y_h}h),
    \label{eq:FSS_Z}
\end{align}
with a scaling function $\tilde{Z}$. 
In Appendix~\ref{app:correction}, we generalize the argument to include the effects of irrelevant operators. 
The numerical influence of the dominant irrelevant operator on our results is studied in Sec.~\ref{sec:FSSviolation}.

\subsection{LYZ ratios}
\label{sec:LYZRs}

In Ref.~\citen{Wada:2024qsk}, we proposed a method to determine $\Tc$ and the critical exponents using the FSS of LYZs in general systems. In this method, which we call the LYZ ratio (LYZR) method, we focus on the ratios of the imaginary parts of the LYZs
\begin{align}
    R_{nm}(t,L) = \frac{\textrm{Im}\, h_\textrm{LY}^{(n)}(t,L)}{\textrm{Im}\, h_\textrm{LY}^{(m)}(t,L)} .
    \label{eq:LYZRs}
\end{align}
In the Ising model, due to the Lee-Yang theorem, Eq.~\eqref{eq:LYZRs} can also be written as 
\begin{align}
    R_{nm}(t,L) = \frac{h_\textrm{LY}^{(n)}(t,L)}{h_\textrm{LY}^{(m)}(t,L)}.
    \label{eq:LYZRs2}
\end{align}

\begin{figure}
    \centering
    \includegraphics[width=0.45\linewidth]{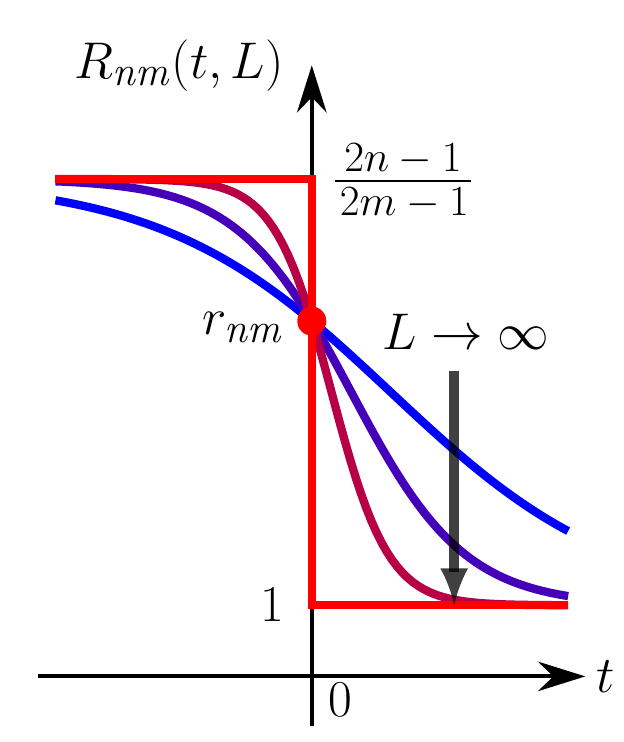}
    \caption{Schmatic behavier of the LYZR $R_{nm}(t,L)$ for $n>m$, as functions of the reduced temperature $t$ for various system sizes $L$. The LYZRs for various $L$ intersect at the CP at $t=0$. The slope at the CP becomes steeper as $L$ increases. In the $L\rightarrow\infty$ limit, $R_{nm}(t,L)$ behaves as a step function shown by the red line, whose value is $(2n-1)/(2m-1)$ and unity for $t<0$ and $t>0$, respectively.} 
    \label{fig:cartoon}
\end{figure}

Let us examine the behavior of the LYZRs in the large volume limit $L\to\infty$. First, for $t<0$ in response to the first-order phase transition at $h=0$ it is known that the LYZs are distributed at equal distances on the imaginary-$h$ axis as
\begin{align}
    h_\textrm{LY}^{(n)}(t,L) \; \xrightarrow[L\to\infty]{} \; a(t)\,\frac{2n-1}{L^3}\qquad  (n \ge 1),
    \label{eq:h_LY_1st}
\end{align}
where $a(t)$ is a purely imaginary function of $t$~\cite{Fisher1965}. The distribution around the real axis becomes dense for $L\to\infty$ in accordance with the discontinuity of $Z(t,h,L)$ at the first-order phase transition in this limit. 
From Eq.~\eqref{eq:h_LY_1st}, one finds that the LYZRs behave as 
\begin{align}
    R_{nm}(t,L) \; \xrightarrow[L\to\infty]{} \;
    \frac{2n-1}{2m-1} \qquad (t<0).
    \label{eq:Rnm_t<0}
\end{align}
Second, for $t>0$, because $Z(t,h,L)$ is analytic at $h=0$, the LYZs are not accumulated around the real axis for $L\to\infty$. Instead, the LYZs in this limit are known to distribute continuously for $|{\rm Im}h|>h_{\rm LYES}(t)$, where the edges of the distribution $h=\pm i h_{\rm LYES}(t)$ is called the Lee-Yang edge singularity~\cite{Kortman:1971zz,bena:2005}. As a result, the LYZRs behave as 
\begin{align}
    R_{nm}(t,L) \; \xrightarrow[L\to\infty]{} \; 1 \qquad (t>0),
    \label{eq:Rnm_t>0}
\end{align}
for finite $n,m$.
Equations~\eqref{eq:Rnm_t<0} and~\eqref{eq:Rnm_t>0} show that the values of $R_{nm}(t,L)$ suddenly change at the CP at $t=0$ as depicted in Fig.~\ref{fig:cartoon} by the red-solid line. This property is similar to that of the Binder cumulant~\cite{Binder:1981sa} that will be discussed in the next subsection.

Next, let us inspect the behavior of $R_{nm}(t,L)$ near the CP at finite $L$. Assuming that the $n$th LYZ is a zero of $Z_{\rm sing}(t,h,L)$, it is shown from Eq.~\eqref{eq:FSS_Z} that $h_\textrm{LY}^{(n)}(t,L)$ satisfies
\begin{align}
    Z_{\rm sing}\left(t, h_\textrm{LY}^{(n)}(t,L),L\right) 
    = \tilde Z\left(L^{y_t}t , L^{y_h} h_\textrm{LY}^{(n)}(t,L)\right) 
    = 0.
\end{align}
Then, by defining zeros of the scaling function as $\tilde h=\tilde{h}_{\rm LY}^{(n)}(\tilde t\,)$, i.e. 
$\tilde Z(\tilde t,\tilde{h}_{\rm LY}^{(n)}(\tilde t\,))=0$, the $n$th LYZ for different $L$ obey~\cite{Itzykson:1983gb}
\begin{align}
    L^{y_h} h_{\rm LY}^{(n)}(t,L) = \tilde{h}_{\rm LY}^{(n)}(L^{y_t}t) .
    \label{eq:tildeh_LY}
\end{align}

As discussed in Sec.~\ref{sec:LYZ}, the LYZs $h_\textrm{LY}^{(n)}(t,L)$, and hence $\tilde{h}_{\rm LY}^{(n)}(\tilde t\,)$, are analytic functions of $t$ and $\tilde t$, respectively. Therefore, $\tilde{h}_{\rm LY}^{(n)}(\tilde t\,)$ are Taylor expanded at $\tilde t=0$ as
\begin{align}
    \tilde{h}_\textrm{LY}^{(n)}(\tilde{t}\,) 
    = i \, \Big(X_n+Y_n \tilde{t} + Z_n \tilde{t}\,^2
    +{\cal O}(\tilde{t}\,^3)\Big) ,
    \label{eq:expansion}
\end{align}
where $X_n, Y_n$, and $Z_n$ are real coefficients. Combining Eqs.~\eqref{eq:expansion} and~\eqref{eq:tildeh_LY}, one finds~\cite{Wada:2024qsk} 
\begin{align}
    R_{nm}(t,L) 
    = r_{nm} + c_{nm} L^{y_t}t + d_{nm} (L^{y_t}t)^2 +{\cal O}(t^3) ,
    \label{eq:LYZRs_expansion}
\end{align}
with $r_{nm}=X_n/X_m$ and $c_{nm} = r_{nm}(Y_n/X_n -Y_m/X_m)$. The coefficients $d_{nm}$ are also represented by the coefficients in Eq.~\eqref{eq:expansion}.

Equation~\eqref{eq:LYZRs_expansion} shows notable features of $R_{nm}(t,L)$ around $t=0$. At the CP, the LYZRs satisfy $r_{nm}=R_{nm}(0,L)$, which is independent of $L$, while the slope scales as $L^{y_t}$. This means that $R_{nm}(t,L)$ obtained at various $L$'s intersect at a single point at $t=0$ as shown in Fig.~\ref{fig:cartoon}.
Therefore, the CP can be located from the intersection point of $R_{nm}(0,L)$ for various $L$.

The extension of this method to the CPs in general systems is discussed in Ref.~\citen{Wada:2024qsk}, where it has been shown that the values of the LYZRs at the CP, $r_{nm}$, are universal constants specific to individual universality classes. 

\subsection{Binder cumulant}
\label{sec:Binder}

Conventional quantities used to identify the CP are the cumulants of the magnetization $M$. From the free energy $F=-T\ln Z$, the cumulants are given by~\cite{Asakawa:2015ybt} 
\begin{align}
    \langle M^n(t,h,L)\rangle_{\rm c} = - T^{n-1}\,\frac{\partial^n F(t,h,L)}{\partial h^n} \bigg|_{T}.
    \label{eq:def_cumulant}
\end{align}
By taking only the singular part $F_{\rm sing}(t,h,L)$ for the free energy assuming sufficiently small $t$ and large $L$, and combining Eq.~\eqref{eq:def_cumulant} with Eq.~\eqref{eq:FSS_F} one obtains
\begin{align}
    \langle M^n(t,h,L)\rangle_{\rm c} = -L^{ny_h} \tilde{F}^{(n)}(L^{y_t}t,L^{y_h}h),
    \label{eq:cumulant_magnetization}
\end{align}
with $\tilde{F}^{(n)}=T^{n-1}\partial^n\tilde{F}/\partial h^n$.
For $h=0$, Eq.~\eqref{eq:cumulant_magnetization} is written as 
\begin{align}
    \langle M^n(t,0,L)\rangle_{\textrm{c}} = - L^{ny_h}\tilde{F}^{(n)}(L^{y_t}t,0).
    \label{eq:Mn_h=0}
\end{align}

To make systematic use of Eq.~\eqref{eq:Mn_h=0}, it is convenient to consider ratios of the cumulants. For example, the ratio of the fourth-order cumulant and the square of the second-order cumulant is called the fourth-order Binder cumulant~\cite{Binder:1981sa}. In the present study we introduce 
\begin{align}
    B_4(t,L) = \frac{\langle{M^4(t,0,L)}\rangle_{\rm c}}{\langle{M^2(t,0,L)}\rangle^2_{\rm c}} + 3 = \frac{\tilde{F}^{(4)}(L^{y_t}t,0)}{\left(\tilde{F}^{(2)}(L^{y_t}t,0)\right)^2} + 3 ,
    \label{eq:B4def}
\end{align}
and refer to it as the Binder cumulant.
These ratios share similar scaling properties to the LYZRs. First, Eq.~\eqref{eq:B4def} in the $L\to\infty$ limit behaves as 
\begin{align}
    B_4(t,L) \xrightarrow[L\to\infty]{}
    \begin{cases}
        3 & (t>0) \\
        1 & (t<0)
    \end{cases}
    \qquad
    (\mbox{finite}~ n,m).
    \label{eq:B4lim}
\end{align}
Equation~\eqref{eq:B4lim} for $t<0$ is obtained from the fact that the system is in a coexisting phase and the distribution of $M$ is two-peaked, while that for $t>0$ is a consequence of the Gauss distribution of $M$. 

Second, near the CP at $t=0$, substituting the Taylor expansion of $\tilde F^{(n)}(\tilde t,0)$
\begin{align}
    \tilde F^{(n)}(\tilde t,0) = \tilde F^{(n)}_0 + \tilde F^{(n)}_1 \tilde t + \frac12 \tilde F^{(n)}_2 \tilde{t}^{\,2} + \cdots,
\end{align}
into Eq.~\eqref{eq:B4def}, one obtains  
\begin{align}
    B_4(t,L) = b_4 + c_4L^{y_t}t + d_4(L^{y_t}t)^2 + {\cal O}(t^3),
    \label{eq:B4_expansion}
\end{align}
with $b_4=\tilde F^{(4)}_0/(\tilde F^{(2)}_0)^2 +3$, $ c_4 = (F_1^{(4)}F_0^{(2)}-2F_0^{(4)}F_1^{(2)})/(F_0^{(2)})^3$, and so forth.
Equations~\eqref{eq:B4_expansion} shows that $B_4(t,L)$ shares properties similar to those of the LYZRs. It takes a unique value $B_4(0,L)=b_4$ for any $L$ at $t=0$, while the $t$ derivative scales as $L^{y_t}$. Therefore, the intersection point of $B_4(t,L)$ for various $L$ determines the location of the CP. The value at the intersection point, $b_4$, is a universal constant characteristic of the universality class~\cite{Binder:1981sa}. In the 3d-Ising model, its value is known as~\cite{Ferrenberg:2018zst} 
\begin{align}
    b_4 = 1.60356(15).
    \label{eq:B4atCP}
\end{align}
One can also readily verify that the ratios of even higher-order cumulants, such as $\langle M^6(t,0,L)\rangle_{\rm c}/\langle M^2(t,0,L)\rangle^3_{\rm c}$, exhibit the same property.

One can further construct quantities having the intersection property at the CP besides the LYZRs and Binder-cumulants~\cite{ Nomura:1995tq, Nomura:1997pp, Tomita_2002}. An example is the one constructed from the correlation function $C(\ell)=\langle \sigma_i \sigma_j \rangle|_{|i-j|=\ell}$~\cite{Tomita_2002}. It is known that its ratio at different distances, $C(2\ell)/C(\ell)$, intersects at $t=0$ and thus can be used for the determination of the CP. However, we do not discuss them further in this paper.

\subsection{Intersection analysis with single LYZ and cumulant}
\label{sec:single_method}

In this subsection, we propose an alternative intersection analysis that uses a single LYZ assuming the known value of the critical exponent $y_h$.

For this purpose, we note that Eqs.~\eqref{eq:tildeh_LY} and \eqref{eq:expansion} lead to
\begin{align}
    L^{y_h} h_\textrm{LY}^{(n)}(t,L) = \tilde{h}_\textrm{LY}^{(n)}(L^{y_t}t) = X_n + Y_n L^{y_t} t + Z_n (L^{y_t} t)^2 + {\cal O}(t^3) ,
    \label{eq:singleLY}
\end{align}
around $t=0$. From Eq.~\eqref{eq:singleLY} one easily finds that $L^{y_h} h_\textrm{LY}^{(n)}(t,L)$ for various $L$ intersect at $t=0$. Therefore, the CP can be identified through intersection analysis of $L^{y_h} h_\textrm{LY}^{(n)}(t,L)$ for different $L$. However, it should be noted that the crossing value $X_n$ is not universal in this case.

This idea is equally applicable to the cumulants of magnetization. A transform of Eq.~\eqref{eq:Mn_h=0}, 
\begin{align}
    \frac{\langle M^n(t,0,L)\rangle_{\rm c}}{L^{ny_h}} = \tilde{F}^{(n)} (L^{y_t}t,0)
    = \tilde{F}^{(0)}_n 
    + \tilde{F}^{(1)}_n L^{y_t}t
    + \tilde{F}^{(2)}_n (L^{y_t}t)^2 
    + {\cal O}(t^3).
    \label{eq:singleM}
\end{align}
immediately tells us that the left-hand side, $\langle M^n(t,0,L)\rangle_{\rm c}/L^{ny_h}$, shares the same property as Eq.~\eqref{eq:singleLY}. 

We refer to these methods as the single-LYZ and single-cumulant methods, respectively. These methods are tested in Sec.~\ref{sec:Single_result}. 


\section{Numerical setup}
\label{sec:setup}

We perform Monte Carlo simulations of the 3d-Ising model~\eqref{eq:def_Ising_H} using the Wolff cluster algorithm~\cite{Swendsen:1987ce,Wolff:1988uh}. 
We fix $h=0$ and generate spin configurations at three temperatures around $\Tc$, $T = 4.51125$, $4.5115$, and $4.51175$, for the lattice sizes $L=16, 24, 32, 48, 64, 96, 128, 192$, and $256$. 
For thermalization, at least $4\times10^5$ Monte Carlo updates are performed starting from random spin configurations. After thermalization, $2\times 10^6$ measurements are taken every ten Monte Carlo steps at each simulation point. 

From the Monte Carlo history of the magnetization~\eqref{eq:M}, we found that the autocorrelation time is shorter than three measurements ($30$ Monte Carlo steps) for $L\le64$ and about $10$ for $L=256$.
We estimate statistical errors using the jackknife method with $20$ bins, corresponding to the binsize of $10^5$ measurements. Since the binsize is several orders of magnitude larger than the autocorrelation time, the effect of autocorrelation is well suppressed in these error analyses.

To search for LYZs on the complex-$h$ plane, we analyze the normalized partition function 
\begin{align}
    N(t,h,L)
    =\frac{Z(t,h,L)}{Z(t,0,L)} ,
    \label{eq:N}
\end{align}
for $h\in\mathbb{C}$. 
Equation~\eqref{eq:N} is evaluated using the reweighting method~\cite{Ferrenberg:1989ui,Ejiri:2005ts} as
\begin{align}
    N(t,h,L)
    =\frac{\langle e^{-H(h,L)/T+H(0,L)/T_0}\rangle_0}{\langle e^{-H(0,L)/T+H(0,L)/T_0}\rangle_0},
    \label{eq:N_reweight}
\end{align}
where $\langle\cdot\rangle_0$ means the expectation value on the spin configurations at a simulation temperature $T=T_0$. 
Since the denominator of Eq.~\eqref{eq:N} never diverges, the LYZs correspond to zeros of Eq.~\eqref{eq:N}. 
To find zeros of Eq.~\eqref{eq:N}, we numerically solve the simultaneous equations
\begin{align}
    \mathrm{Re}\, N(t,h,L) =0
    , \qquad
    \mathrm{Im}\, N(t,h,L) =0 ,
    \label{eq:Nr}
\end{align}
for fixed $t$ and $L$.

For each $L$ and $T$, we determine the LYZ by solving Eq.~\eqref{eq:Nr} at three simulation temperatures $T_0$ independently. Statistical errors are estimated by solving Eq.~\eqref{eq:Nr} on individual jackknife samples. The final result is obtained as the weighted average over the three $T_0$ results.
In the analyses of the LYZRs~\eqref{eq:LYZRs}, the ratios are computed within individual jackknife samples. This procedure is crucial for proper estimates of statistical errors, eliminating the correlations among LYZs. In the Binder-cumulant analysis, we apply the same procedure, computing cumulants and their ratios within individual jackknife samples.

\begin{figure}
        \centering
        \includegraphics[width=0.98\linewidth]{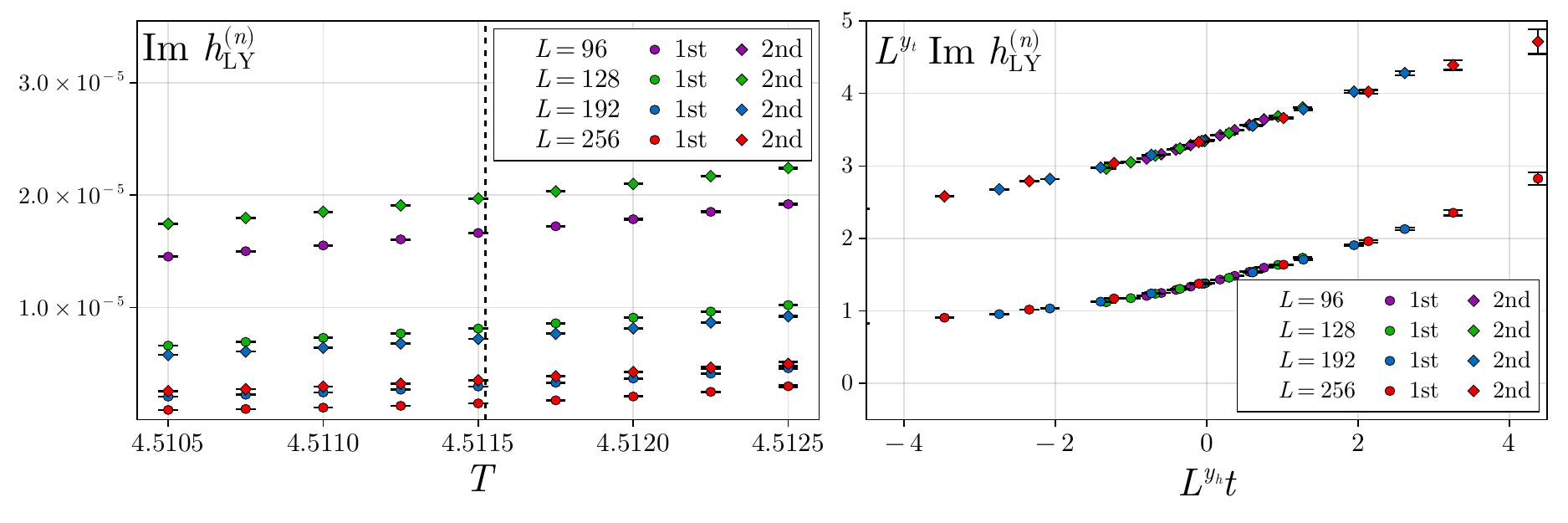}
        \caption{
        Left: Imaginary parts of the first and second LYZs, ${\rm Im}\,h_{\rm LY}^{(n)}(T,L)$ with $n=1,2$, for various $L$. The vertical dashed line represents $\Tc$ in Eq.~\eqref{eq:Tc}~\cite{Ferrenberg:2018zst}. Right: the same data with the vertical and horizontal axes rescaled according to the scaling relation Eq.~\eqref{eq:tildeh_LY} with $t=(T-T_{\rm c})/T_{\rm c}$.}
        \label{fig:LYZs}
\end{figure}

The left panel of Fig.~\ref{fig:LYZs} shows the imaginary parts of the first and second LYZs, ${\rm Im}\, h_{\rm LY}^{(1)}(T,L)$ and ${\rm Im}\, h_{\rm LY}^{(2)}(T,L)$, for $L=96$, 128, 192, and 256 as functions of $T$ around $\Tc$.
The first (second) LYZ for each $L$ is shown by the circle (diamond) symbols. The statistical errors are depicted but much smaller than the symbols and are almost not visible. The vertical dashed line represents the location of $\Tc$ in Eq.~\eqref{eq:Tc}. 

The right panel of Fig.~\ref{fig:LYZs} shows the same results, with both the vertical and horizontal axes rescaled according to the scaling relation Eq.~\eqref{eq:tildeh_LY}, using Eqs.~\eqref{eq:ytyh} and~\eqref{eq:Tc} for critical exponents and $\Tc$, respectively.
The figure shows that the first and second LYZs for various $L$ accumulate on individual universal curves, as expected from Eq.~\eqref{eq:tildeh_LY}. We confirmed that the third and fourth LYZs exhibit the same scaling behavior.

\section{Lee-Yang-zero ratios and Binder cumulant}
\label{sec:Numerical_result}

\subsection{Intersection analysis}
\label{sec:intersection}

\begin{figure}
    \centering
    \includegraphics[width=0.98\linewidth]{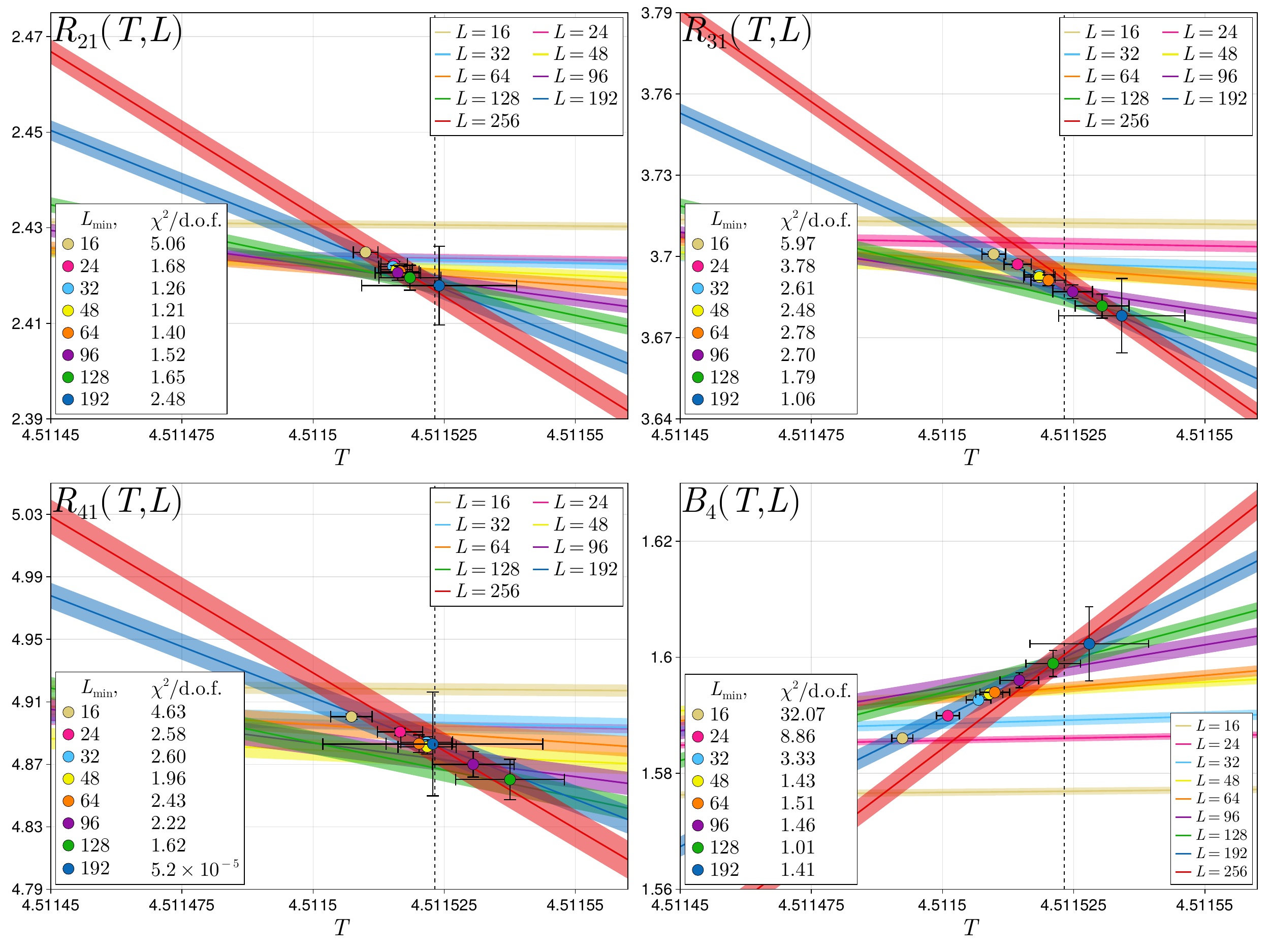}
    \caption{LYZRs $R_{n1}(T,L)$ for $n=2,3,4$ and the fourth-order Binder-cumulant $B_4(T,L)$ for various $L$. Statistical errors are indicated by the shaded bands. The circle markers with error bands show the fit results for various smallest system size $L_{\rm min}$ with $\Delta T \times 10^5 = 5$.}
    \label{fig:LYZRs_L_fit}
\end{figure}

In Fig.~\ref{fig:LYZRs_L_fit}, we present the behavior of LYZRs $R_{n1}(T,L)$ for $n=2,3,4$, as well as the fourth-order Binder cumulant $B_4(T,L)$, as functions of $T$ for various system size $L$.
The shaded bands indicate statistical errors. In this study, we consider only $R_{21}(T,L)$, $R_{31}(T,L)$, $R_{41}(T,L)$ for the LYZRs, since the statistical errors of $R_{nm}(T,L)$ typically grow with increasing $n$ and $m$~\cite{Wada:2024qsk}. 
The figure shows that $R_{n1}(t,L)$ and $B_4(t,L)$ obtained at various $L$ intersect at almost a common point, as anticipated from Eqs.~\eqref{eq:LYZRs_expansion} and~\eqref{eq:B4_expansion}. However, results for smaller $L$ clearly deviate from this trend. The deviation at small $L$ is interpreted as a consequence of the correction to FSS. Comparing the four panels, the convergence toward the $L\to\infty$ limit is the slowest in $B_4(t,L)$, suggesting that this quantity exhibits the strongest correction to FSS. We will come back to this point in Sec.~\ref{sec:FSSviolation}.
In Fig.~\ref{fig:LYZRs_L_fit}, the value of $T_{\rm c}$ in Eq.~\eqref{eq:Tc} is indicated by the vertical dashed lines. The temperature at the intersection is close to this value in all cases.

To determine the intersection point quantitatively, we perform the $\chi^2$ fits to these data. 
We use the following fitting function for the LYZRs motivated by Eq.~\eqref{eq:LYZRs_expansion}
\begin{align}
    R_{nm}(T,L)=
        r_{nm} + c_{nm}\,L^{y_t}\,\bigg(\frac{T-\Tc}{\Tc}\bigg) + d_{nm}\,L^{2y_t}\,\bigg(\frac{T-\Tc}{\Tc}\bigg)^2 ,
        \label{eq:LYZR_nonlinear}
\end{align}
where $r_{nm}$, $c_{nm}$, $d_{nm}$, $y_t$, and $\Tc$ are the fitting parameters. 
For $B_4(T,L)$, we use the same functional form
\begin{align}
    B_4(T,L)=
        b_4 + c_4\,L^{y_t}\,\bigg(\frac{T-\Tc}{\Tc}\bigg) + d_4\,L^{2y_t}\,\bigg(\frac{T-\Tc}{\Tc}\bigg)^2,
        \label{eq:Binder_nonlinear}
\end{align}
with $b_4$, $c_4$, $d_4$, $y_t$, and $\Tc$ the fitting parameters. 

In our analysis that employs the reweighting method, the $T$ values at which the data for $R_{n1}(T,L)$ or $B_4(T,L)$ are referenced can be chosen arbitrarily. In this study, we use three $T$ values for each $L$ for the fits. These values should be close to $T_{\rm c}$ to suppress the nonsingular effects. However, they should be separated so that the correlations arising from the reweighting method on the same configurations are not too large. Following the analysis in Appendix~\ref{app:DeltaT}, for the three $T$ values we choose $T=4.5115$ and $4.5115\pm\Delta T$ with $\Delta T=5\times10^{-5}$. 
As discussed in Appendix~\ref{app:DeltaT}, this $\Delta T$ is chosen so that the fit results are insensitive to the variation of $\Delta T$ around this value. To account for the correlations between different $T$ values, we perform correlated $\chi^2$ fits, which take into account the correlation between the data points by the covariance matrix.

We notice that the nonlinear term proportional to $d_{n1}$ or $d_4$ in Eqs.~\eqref{eq:LYZR_nonlinear} and~\eqref{eq:Binder_nonlinear} is necessary for these fitting analyses. In fact, we have checked that the four-parameter fits with $d_{n1}=0$ or $d_4=0$ give unacceptably large $\chi^2/{\rm d.o.f.}$ except for $R_{41}(T,L)$. 
The magnitude of this non-linearity will be discussed in more detail in Sec.~\ref{sec:nonlinear}.

\begin{table}[t]
    \centering
    \caption{Results of the fit parameters $T_{\rm c}$, $y_t$, and $r_{n1}$ or $b_4$ obtained by the intersection analysis of the LYZR and Binder-cumulant methods for various smallest system size $L_\textrm{min}$.} 
    \begin{tabular}{cc|lll}\hline
         & $L_{\rm min}$ & $\Tc$ & $y_t$ & $r_{n1}$ or $b_4$ \\ \hline
         $R_{21}$ 
         & 192 & 4.5115240(148) & 1.5338(410) & 2.4179(82)\\
         & 128 & 4.5115185(59) & 1.5874(115) & 2.4195(26)\\
         & 96 &  4.5115161(43) & 1.5947(72) & 2.4206(15)\\
         & 64 &  4.5115166(36) & 1.5891(56) & 2.4208(11)\\
         & 48 &  4.5115159(30) & 1.5890(32) & 2.4213(08)\\
         \hline
         $R_{31}$
         & 192 & 4.5115342(120) & 1.4899(375) & 3.6781(138) \\
         & 128 & 4.5115304(51) & 1.5662(121) & 3.6817(45) \\
         & 96  & 4.5115248(38) & 1.5880(72) & 3.6870(25) \\
         & 64  & 4.5115202(33) & 1.5885(53) & 3.6912(20) \\
         & 48  & 4.5115185(30) & 1.5893(31) & 3.6924(16)\\
         \hline
         $R_{41}$
         & 192 & 4.5115228(21) & 1.4834(475) & 4.8830(332)  \\
         & 128 & 4.5115375(104)& 1.5311(185) & 4.8603(130)  \\
         & 96  & 4.5115305(78) & 1.5680(113) & 4.8700(82)  \\
         & 64  & 4.5115202(63) & 1.5844(84)  & 4.8832(56)  \\
         & 48  & 4.5115217(56) & 1.5841(51)  & 4.8812(44) \\
         \hline
         $B_4$
         & 192 & 4.5115280(113)& 1.5313(267)& 1.60230(638)\\
         & 128 & 4.5115211(52) & 1.5691(86) & 1.59892(224)\\ 
         & 96  & 4.5115146(37) & 1.5829(52) & 1.59604(130)\\
         & 64  & 4.5115099(28) & 1.5888(33) & 1.59397(72)\\
         & 48  & 4.5115089(26) & 1.5880(21) & 1.59368(56)\\\hline
    \end{tabular}
    \label{tab:Final_Result_1}
\end{table}

In Fig.~\ref{fig:LYZRs_L_fit}, we show the fit results for the intersection point obtained by the five-parameter correlated fits with Eqs.~\eqref{eq:LYZR_nonlinear} and~\eqref{eq:Binder_nonlinear} by the colored symbols. To estimate the magnitude of the correction to FSS, the fits are performed for various smallest system sizes $L_\textrm{min}$ used for the fits, while the largest size is fixed to $L=256$. For the largest $L_{\rm min}=192$, the number of degrees of freedom (d.o.f.) is one. The values of $\chi^2/{\rm d.o.f.}$ are shown in the figure, and the resulting values of $T_{\rm c}$, $y_t$, and $r_{n1}$ or $b_4$ are shown in Table~\ref{tab:Final_Result_1} for $L_{\rm min}\ge48$.

\begin{table}[t]
    \centering
    \caption{Values of LYZRs at the CP $r_{n1}$. The upper row shows the results obtained by the intersection analysis with $L_{\rm min}=128$. Shown in the middle row are the values obtained by the extrapolation to $L\to\infty$ via Eq.~\eqref{eq:largeV} at $T=T_{\rm c}$  in Eq.~\eqref{eq:Tc}. Systematic uncertainties of the extrapolation are denoted by the second error. The values obtained for the CP in the three-dimensional three-state Potts model~\cite{Wada:2024qsk} are shown in the lower panel. } 
    \begin{tabular}{l|lll}
    \hline
    & $r_{21}$ & $r_{31}$ & $r_{41}$ \\
    \hline
    intersection analysis ($L_{\rm min}=128$)
    & $2.4195(26)$ & $3.6817(45)$ & $4.8603(130)$ \\
    extrapolation at $T=T_{\rm c}$ in Eq.~\eqref{eq:Tc}
    & $\mathbf{2.4158}(37)(5)$ & $\mathbf{3.6906}(68)(89)$ & $\mathbf{4.8896}(156)(23)$ \\
    3-state Potts model~\cite{Wada:2024qsk}    & $2.408(12)$ & $3.669(24)$ & $4.861(36)$ \\ \hline
    \end{tabular}
    \label{tab:LYZR_final}
\end{table}

These results show that the intersection points converge well within statistics for $L_\textrm{min} \ge 24$, although the convergence of $B_4(T,L)$ may be slightly slower than $R_{n1}(T,L)$. The statistical uncertainties of $T_{\rm c}$ in these results are comparable at the same $L_{\rm min}$, whereas those of $R_{41}(T,L)$ are clearly larger than the others. The obtained values of $T_{\rm c}$, $y_t$, and $b_4$ are all consistent with the known results, Eqs.~\eqref{eq:Tc}, \eqref{eq:ytyh}, and \eqref{eq:B4atCP}, within about twice the standard deviation. From these results, we conclude that the LYZR method can determine $T_{\rm c}$ and $y_t$ in the Ising model with nearly the same precision as the Binder-cumulant method.

In the following, we use the fit results at $L_{\rm min}=128$ for the final result of the intersection analysis.
The resulting values of $r_{n1}$ are shown on the top row of Table~\ref{tab:LYZR_final}.
In Ref.~\citen{Wada:2024qsk}, we studied the CP in the three-dimensional three-state Potts model with an external field using the LYZR method. This CP is believed to belong to the same universality class as the 3d-Ising model~\cite{Karsch:2000xv,Falcone:2006na}. As shown in Ref.~\citen{Wada:2024qsk}, the values of $r_{nm}$ in the same universality class should be the same. The values of $r_{nm}$ at the intersection point obtained in Ref.~\citen{Wada:2024qsk} are
shown on the bottom row in the table. The table shows that they are consistent with each other. The statistical errors, however, are significantly suppressed in the 3d-Ising model.

\begin{figure}
    \centering
    \includegraphics[width=0.95\linewidth]{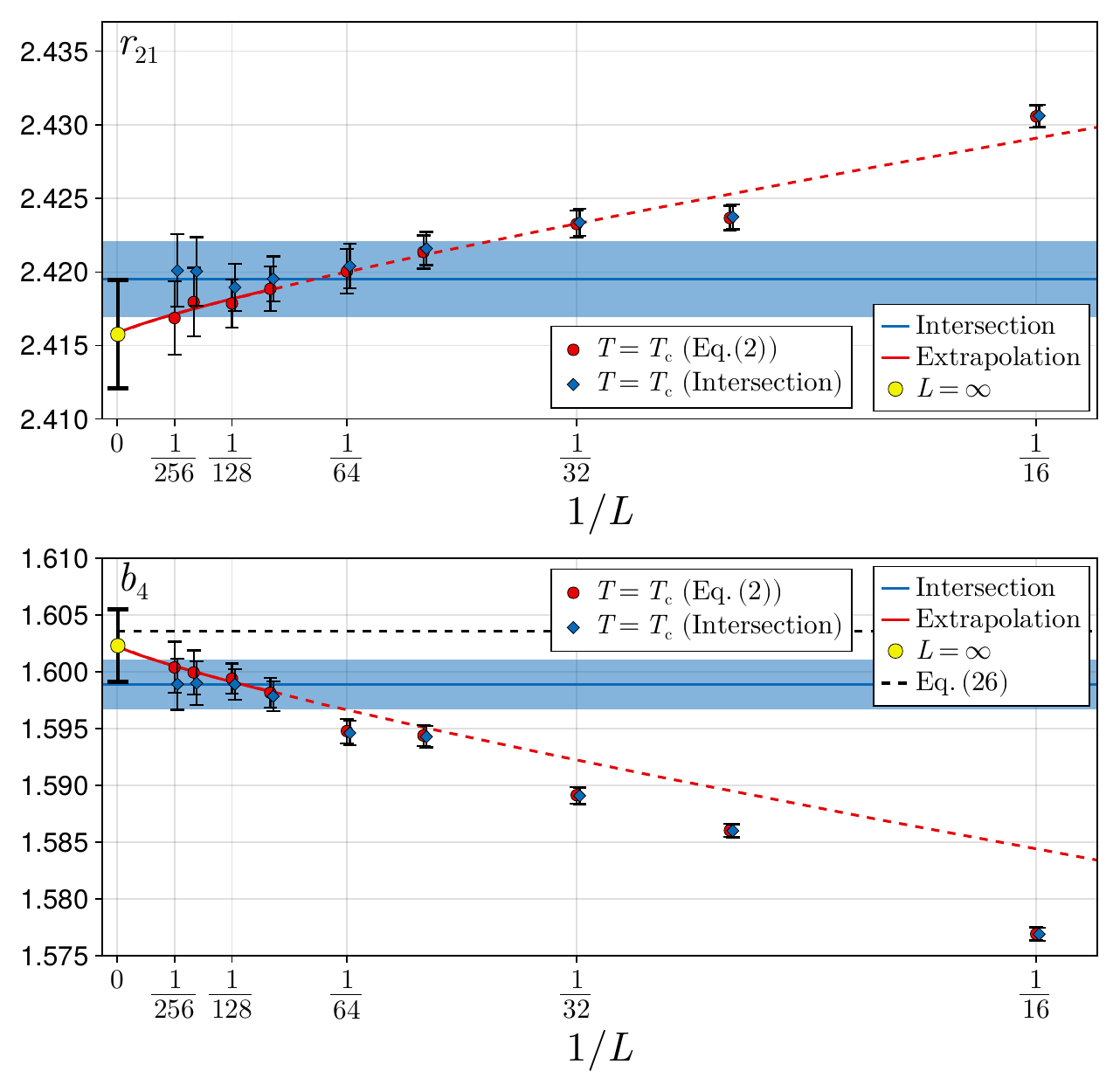}
    \caption{LYZR and fourth-order Binder cumulant at $T=T_{\rm c}$, $R_{21}(T_{\rm c},L)$ and $B_4(T_{\rm c},L)$, as functions of $1/L$. The blue diamonds represent the results at $T=\Tc$ evaluated in the intersection analysis with $L_{\rm min}=128$, while the red circles represent those at Eq.~\eqref{eq:Tc}. The horizontal solid lines are the values of $r_{21}$ and $b_{4}$ obtained by the intersection analysis at $L_{\rm min}=128$ together with the statistical errors depicted by the color bands. The horizontal black-dashed line in the lower panel denotes $b_4$ in Eq.~\eqref{eq:B4atCP}. The red solid/dashed lines are the fit result with Eq.~\eqref{eq:largeV} for the data at $L\ge96$, and the yellow symbols are the values of $r_{21}$ and $b_4$ obtained by the $L\to\infty$ extrapolations.}
    \label{fig:LYZRs_B4}
\end{figure}

\subsection{correction to FSS}
\label{sec:FSSviolation}

The fit analyses based on Eqs.~\eqref{eq:LYZR_nonlinear} and~\eqref{eq:Binder_nonlinear} in the previous subsection assume the validity of the FSS relations~\eqref{eq:FSS_F} or~\eqref{eq:FSS_Z}. As Fig.~\ref{fig:LYZRs_L_fit} shows, however, our numerical results at small $L$ clearly do not intersect at the same point, indicating the correction to the scaling. 
In this section, we investigate this effect in more detail and perform the analysis of $r_{n1}$ after eliminating its effect.

In Fig.~\ref{fig:LYZRs_B4}, we show the values of $R_{21}(T,L)$ and $B_4(T,L)$ at $T=T_{\rm c}$ determined by the intersection analysis based on Eqs.~\eqref{eq:LYZR_nonlinear} and~\eqref{eq:Binder_nonlinear} at $L_{\rm min}=128$ by the blue diamonds as functions of $1/L$. For comparison, we also plot the same quantities at $T=T_{\rm c}$ in Eq.~\eqref{eq:Tc} by the red circles. Provided that the FSS is valid and $T_{\rm c}$ is determined exactly, these quantities should be independent of $1/L$, $R_{21}(T_{\rm c},L)=r_{21}$ and $B_4(T_{\rm c},L)=b_4$, whose values obtained by the intersection analyses in the previous subsection are shown by the horizontal solid lines in the figure. The black dashed line in the lower panel indicates the previous result of $b_4$ in Eq.~\eqref{eq:B4atCP}. 

We find that the values of $R_{21}(T_{\rm c},L)$ and $B_4(T_{\rm c},L)$ in the figure are consistent with each other for $L\ge96$. However, a clear $L$ dependence is observed in both results for $L\lesssim32$, which can be attributed to the correction to FSS. The comparison of two panels indicates that the correction to FSS is stronger in $B_4(T,L)$ than $R_{21}(T,L)$. This implies that the LYZRs are more effective in suppressing the effect of the correction in the intersection analysis.
The red circles may also suggest a remaining weak dependence on $L$ even for $L\ge96$.

To extract the value of $r_{21}$ while eliminating the effects of the correction to FSS, we perform an extrapolation to the $L\to\infty$ limit.
In Appendix~\ref{app:correction}, we investigate the form of corrections to Eq.~\eqref{eq:LYZRs_expansion} from irrelevant operators. 
From Eq.~\eqref{eqapp:rnm}, we adopt an ansatz for the $L\to\infty$ extrapolation~\cite{Ferrenberg:2018zst}
\begin{align}
    R_{21}(T_{\rm c},L) = r_{21} \, (1 +  cL^{-\omega_1}),
    \label{eq:largeV}
\end{align}
at $T=\Tc$ in Eq.~\eqref{eq:Tc}, where $r_{21}$ and $c$ are the fit parameters. 
The second term represents the correction due to the dominant irrelevant operator. We use the exponent $\omega_1=0.8303(18)$ evaluated in Ref.~\citen{El-Showk:2014dwa} by the conformal bootstrap method. The same fit is also applied to $B_4(T_{\rm c},L)$.
The red curves in Fig.~\ref{fig:LYZRs_B4} represent the results of the fit to $R_{21}(T_{\rm c},L)$ and $B_4(T_{\rm c},L)$ for four system sizes $L$ in the range $96\le L\le256$.
The yellow circles indicate the extrapolated values of $r_{21}$ and $b_4$. The value of $r_{21}$ is shown on the middle row of Table~\ref{tab:LYZR_final}, together with the results of the same analyses for $R_{31}(T_{\rm c},L)$ and $R_{41}(T_{\rm c},L)$. 
The table shows that the extrapolated values of $r_{n1}$ are consistent with those obtained by the intersection analyses within statistics, while the extrapolated values would be more reliable, as they take account of the scaling violation.

To estimate the systematic uncertainties of the extrapolation with Eq.~\eqref{eq:largeV}, we also perform the fits with the variation of $\omega_1$ in the range $0.73<\omega_1<0.93$ and $L_{\rm min}=64, 128$. We take the largest deviation among these analyses as the systematic error of each result, shown as the second error in Table~\ref{tab:LYZR_final}. From the table, one sees that the systematic errors are at most comparable to the statistical errors.

\subsection{Non-linearity}
\label{sec:nonlinear}

In the analysis in Sec.~\ref{sec:intersection}, we found that the non-linearity of $R_{n1}(T,L)$ and $B_4(T,L)$ are not negligible in the range of $T$ used in the fit analyses. Such non-linearity prevents the use of linear fits with $d_{nm}=0$ or $d_4=0$, which are much simpler and usually employed in the literature~\cite{Binder:2001ha,Pelissetto:2000ek,Ferrenberg:2018zst}. To safely apply linear fits, one must choose a narrower temperature range near $T_{\rm c}$ to suppress the nonlinear effects, which, however, typically increases the statistical errors. In this subsection, we compare the magnitude of non-linearity in different methods. Throughout this subsection, we assume the validity of the FSS and neglect its correction for simplicity.

\begin{figure}
    \centering
    \includegraphics[width=0.5\linewidth]{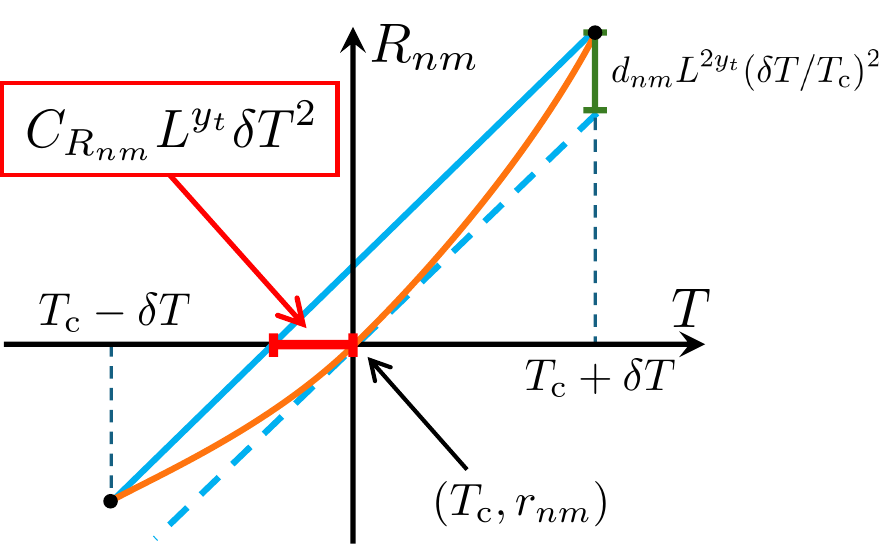}
    \caption{
    Schematic picture of a geometric 
    interpretation of Eq.~\eqref{eq:C_f}. The red band indicates the deviation of $T_{\rm c}$ from the true value caused by 
    the linear approximation in Eq.
    ~\eqref{eq:Rlinear}, which is proportional to $C_{R_{nm}}\delta T^2$.
    }
    \label{fig:nonlinear}
\end{figure}

As a quantitative measure to compare the magnitude of the non-linearity in different methods, we introduce a normalized curvature 
\begin{align}
    C_f = L^{-y_t} \frac{\partial^2 f/\partial T^2}{\partial f/\partial T }\Bigg|_{T=\Tc},
    \label{eq:C_f}
\end{align} 
with $f = R_{nm}(T,L)$ or $B_4(T,L)$.
The meaning of this quantity is understood as follows. First, if one knows the exact value of $r_{nm}$, the critical temperature $T_{\rm c}$ is determined by solving $R_{nm}(T_{\rm c},L)=r_{nm}$. Second, if one evaluates $R_{nm}(T,L)$ at $T=T_{\rm c}\pm \delta T$, an approximate formula of $R_{nm}(T,L)$ by a linear interpolation is given by
\begin{align}
    R_{nm}^{\rm linear}(T,L) 
    = R_{nm}(T_{\rm c}+\delta T,L) \frac{T-T_{\rm c}+\delta T}{2\delta T} - R_{nm}(T_{\rm c}-\delta T,L) \frac{T-T_{\rm c}-\delta T}{2\delta T}.
    \label{eq:Rlinear}
\end{align}
Finally, the solution of $R_{nm}^{\rm linear}(T_{\rm c},L)=r_{nm}$, i.e.\ an estimate of $T_{\rm c}$ based on the linear interpolation, is given by $T_{\rm c}-C_{R_{nm}}L^{y_t} \delta T^2$, as shown by the red band in Fig.~\ref{fig:nonlinear}. Therefore, $C_{R_{nm}}$ represents the deviation of the estimate of $T_{\rm c}$ in the linear interpolation with fixed $\delta T$. Using Eq.~\eqref{eq:LYZRs_expansion} or~\eqref{eq:LYZR_nonlinear}, Eq.~\eqref{eq:C_f} is given by $C_{R_{nm}}=d_{nm}/\Tc c_{nm}$. The same argument also applies to $C_{B_4}$.
The measures $C_f$ are not universal quantities specific to a universality class. However, their ratios, such as $C_{R_{21}}/C_{B_4}$, in the $L\to\infty$ limit are invariant under the nonlinear variable transformations of $T$ and $h$, as discussed in Appendix~\ref{app:second}. In this sense, they can be used for measures to compare the magnitude of non-linearity in different methods. 

\begin{table}[]
    \centering
    \caption{Normalized curvatures $C_f$ for the LYZR $R_{nm}(T,L)$ and the Binder cumulant $B_4(T,L)$.}
    \begin{tabular}{c|cccc}
    \hline
    $f$ & $R_{21}$ & $R_{31}$ & $R_{41}$ & 
    $B_4$ \\
    \hline
    $C_f$ & 0.0093(17) & 0.0053(17) & -0.0010(34) & 0.0364(37) \\ \hline
    \end{tabular}
    \label{tab:C}
\end{table}

In Table~\ref{tab:C}, we show the values of $C_f$ obtained by the nonlinear fits with $L_{\rm min}=128$ and $\Delta T=5\times10^{-5}$. One finds that the value of $C_{B_4}$ is about four times larger than $C_{R_{21}}$, meaning the larger non-linearity in the Binder-cumulant method. In the LYZR method, $C_{R_{n1}}$ for larger $n$ is smaller for $n=2,3,4$. In particular, $C_{R_{41}}$ vanishes within statistics. These results, possibly due to accidental cancellations of the nonlinear terms in the LYZs, 
might be useful in some numerical analyses. It is also notable that different combinations of $n$ and $m$ have different characteristics in the LYZR method. For example,
$R_{41}(T,L)$ greatly suppresses the nonlinear effect, while it has larger statistical errors than $R_{21}(T,L)$ and $R_{31}(T,L)$ as we have seen in Sec.~\ref{sec:intersection}. 

\section{Single-LYZ and single-cumulant methods}\label{sec:Single_result}

\begin{figure}
    \centering
    \includegraphics[width=0.98\linewidth]{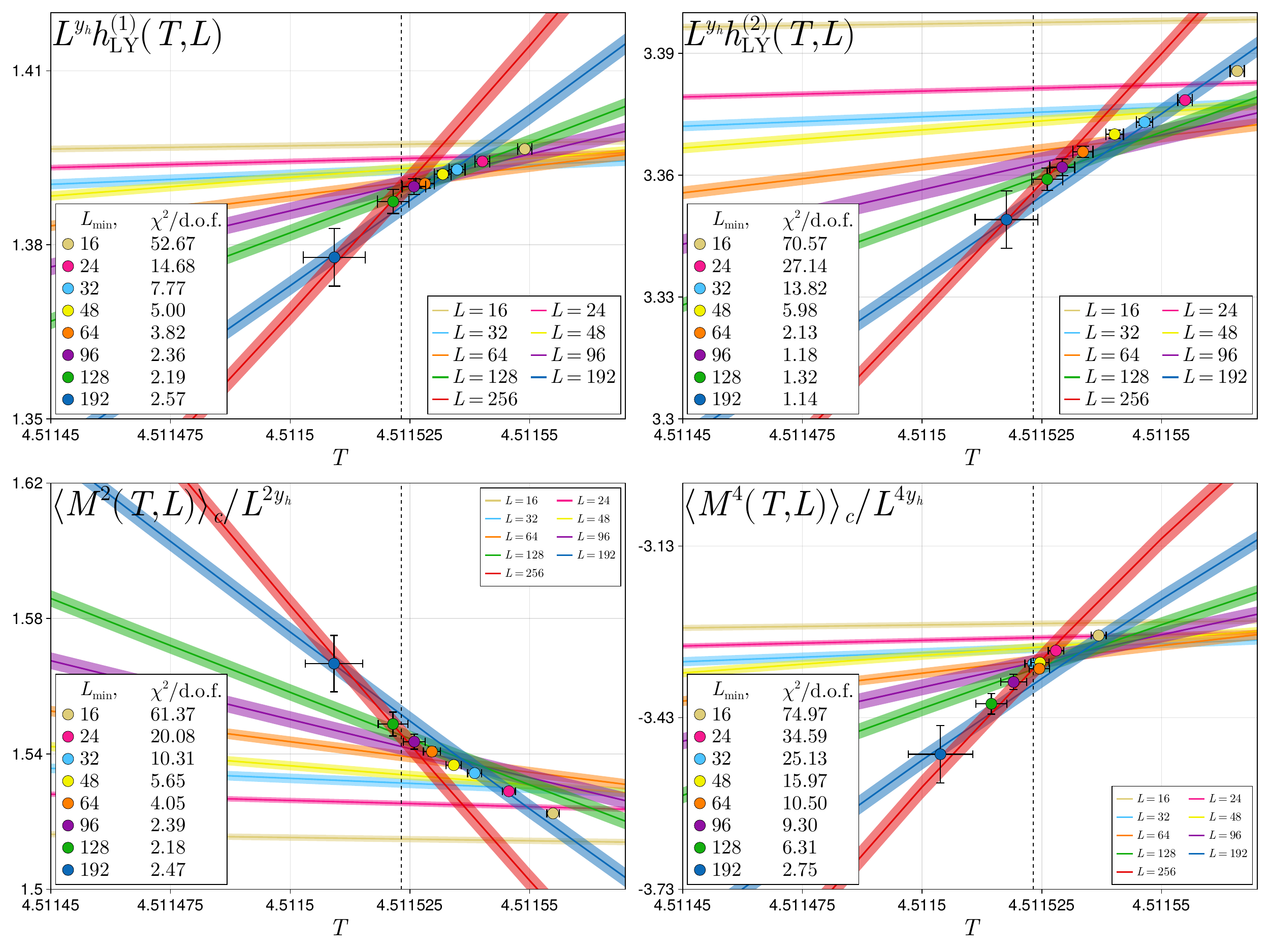}
    \caption{Rescaled LYZs $L^{y_h} h^{(n)}_\textrm{LY}$ for $n=1,2$ and rescaled cumulants $\langle M^n (T,0,L)\rangle_{\rm c}/L^{ny_h}$ for $n=2,4$ at various $L$. The colored circles are the fit results with various $L_\textrm{min}$.}
    \label{fig:LYZs_cumulant_L_fit}
\end{figure}

We now examine the single-LYZ and single-cumulant methods introduced in Sec.~\ref{sec:single_method}. Figure~\ref{fig:LYZs_cumulant_L_fit} peresents the rescaled LYZs,
$L^{y_h} h_\textrm{LY}^{(n)}(T,L)$ for $n=1,2$, and rescaled cumulants, $\langle M^n(T,0,L) \rangle_\textrm{c}/L^{ny_h}$ for $n=2,4$, as functions of $T$, where we use the value of $y_h$ in Eq.~\eqref{eq:ytyh}.
The figure shows that these quantities for various $L$ intersect around the $\Tc$ of Eq.~\eqref{eq:Tc} indicated by the vertical dashed lines. However, the deviations of the intersection point at small $L$ are larger than those of the LYZRs and Binder cumulant in Fig.~\ref{fig:LYZRs_L_fit}. 

To obtain the intersection point quantitatively, we fit these data using fitting functions motivated by Eqs.~\eqref{eq:singleLY} and \eqref{eq:singleM},
\begin{align}
    L^{y_h} h_\textrm{LY}^{(n)}(T,L)
    &= X_n + Y_n L^{y_t} \bigg(\frac{T-\Tc}{\Tc}\bigg) + Z_{n}L^{2y_t} \bigg(\frac{T-\Tc}{\Tc}\bigg)^2,
    \label{eq:LYZ_nonlinear}
    \\
    \frac{\langle M^n(T,0,L)\rangle_{\rm c}}{L^{ny_h}}
    &= \tilde{F}^{(0)}_n + \tilde{F}^{(1)}_n L^{y_t} \bigg(\frac{T-\Tc}{\Tc}\bigg) + \tilde{F}^{(2)}_n L^{2y_t}\bigg(\frac{T-\Tc}{\Tc}\bigg)^2 
\label{eq:cumulant_nonlinear}
\end{align}
where $X_{n}$, $Y_{n}$, $Z_n$, $\tilde{F}^{(m)}$ $(m=0,1,2)$, and $\Tc$ are the fitting parameters. The fitting procedure is the same as before, while in this case we fix $y_t$ to the known value in Eq.~\eqref{eq:ytyh}, resulting in four fitting parameters. The dependence of the fit results on $\Delta T$ is discussed in Appendix~\ref{app:DeltaT}.

In Fig.~\ref{fig:LYZs_cumulant_L_fit}, the intersection points obtained from the fits are shown by the colored circles for various $L_\textrm{min}$. 
We find that, even excluding the $L_\textrm{min}=192$ results with large statistics errors,
the dependence on $L_\textrm{min}$ in the single-LYZ and cumulant methods is stronger than that observed in the LYZR and Binder-cumulant methods. 
From the values of $\chi^2/\textrm{d.o.f.}$ indicated in the figure, we also find that the $\chi^2/\textrm{d.o.f.}$ of the intersection analysis increases more rapidly as $L_{\rm min}$ decreases. 
This indicates that the correction to FSS is more pronounced in the single-LYZ and single-cumulant methods. 
Table~\ref{tab:Csingle} lists the normalized curvatures $C_f$ for $f=h_{\rm LY}^{(n)}$ and $\langle M^n\rangle_{\rm c}$. Their values are significantly larger than those in Table~\ref{tab:C}, indicating that the magnitude of non-linearity is greater in these quantities.

\begin{table}[]
    \centering
    \caption{Normalized curvatures $C_f$ for the single-LYZ and cumulant methods, where $f=h_{\rm LY}^{(n)}$ for $n=1,2$ and $\langle M^n\rangle_{\rm c}$ for $n=2,4$.}
    \label{tab:Csingle}
    \begin{tabular}{c|cccc}
    \hline
    $f$ & $h_{\rm LY}^{(1)}$ & $h_{\rm LY}^{(2)}$ & $\langle M^2\rangle_{\rm c}$ & $\langle M^4\rangle_{\rm c}$ \\
    \hline 
    $C_f$ & 0.07149(59) & 0.04230(101) & -0.07209(44) & -0.17974(104)\\ \hline    
    \end{tabular}
\end{table}

From these results, we conclude that the single methods are less suitable for precise analyses, since they cannot sufficiently suppress systematic uncertainties arising from scaling violation and nonlinear terms. Nevertheless, they are convenient for obtaining a rough estimate of the location of a CP whose universality class is already known.


\section{Summary}\label{sec:Discussion}

We performed Monte Carlo simulations of the three-dimensional Ising (3d-Ising) model to apply the recently proposed Lee-Yang-zero ratio (LYZR) method~\cite{Wada:2024qsk} to this system. The LYZR method enables an intersection analysis analogous to the Binder-cumulant method by exploiting the finite-size scaling (FSS) of the Lee-Yang zeros.
We demonstrated that the LYZR method is as powerful as the Binder-cumulant method in the 3d-Ising model. Moreover, we found that both the correction to FSS and the non-linearities are suppressed in the LYZRs compared with the Binder cumulant. 
The values of the LYZRs at the critical point (CP), which are universal constants, were determined with high precision, as summarized in Table~\ref{tab:LYZR_final}. 

We also introduced alternative intersection analyses that use only a LYZ while assuming the critical exponent $y_h$ as an input. These methods have the advantage that the determination of the second LYZ is unnecessary, and are therefore applicable to systems in which such determination is difficult, such as QCD at nonzero chemical potential~\cite{Schmidt:2025ppy,Adam:2025phc}.
We demonstrated that these methods are useful for studying the CP in the Ising model, although they exhibit stronger corrections to FSS and nonlinear effects.

The LYZR method is applicable to numerical searches for CPs in a wide range of systems, such as magnetic transitions in spin models, CPs in QCD at nonzero baryon density~\cite{Schmidt:2025ppy,Adam:2025phc} and in the heavy-quark region~\cite{Cuteri:2020yke,Philipsen:2021qji,Kiyohara:2021smr,Ashikawa:2024njc}. 
It can also be combined with various numerical approaches, such as the tensor-network method, in addition to Monte Carlo simulations.
Extending the applications of the LYZR method to these directions constitutes an interesting subject for future work.


\begin{acknowledgment}
This work was supported in part by JST SPRING, Grant Number JPMJSP2110, JSPS KAKENHI (Nos.~JP22K03593, JP22K03619, JP23H04507, JP24K07049), ISHIZUE 2025 of Kyoto University, and the Center for Gravitational Physics and Quantum Information (CGPQI) at Yukawa Institute for Theoretical Physics. 
\end{acknowledgment}

\appendix

\section{Contribution of irrelevant operators to FSS}
\label{app:correction}

In the main text, we treated the finite-size effects of various quantities near the CP under the assumption that the singular part obeys the finite-size scaling (FSS), Eq.~\eqref{eq:FSS_F} or~\eqref{eq:FSS_Z}. However, these equations neglect contributions from irrelevant operators. When such contributions are not negligible, Eq.~\eqref{eq:FSS_F} is modified as~\cite{Binder:2001ha,Pelissetto:2000ek,Ferrenberg:2018zst}
\begin{align}
    F_\textrm{sing}(t,h,\{\lambda_i\},L^{-1}) 
    &= \tilde F(L^{y_t}t,\, L^{y_h}h,\, \{L^{y_i}\lambda_i\} ) ,
    \label{eq:FSS_Firr}
\end{align}
where $\{\lambda_i\}$ ($i=1,2,\cdots$) denotes the set of irrelevant operators whose scaling exponents $y_i=-\omega_i<0$ are universal constants within the same universality class. 
In this appendix, we discuss the modifications of Eqs.~\eqref{eq:LYZRs_expansion} and~\eqref{eq:B4_expansion} arising from the contribution of the irrelevant operators. 
The obtained results are used in Sec.~\ref{sec:FSSviolation} to study the correction to FSS in the numerical results.
Throughout this appendix, we assume that thermodynamics is dominated by the singular part and neglect the regular part for simplicity. 

We first focus on the Binder cumulant~\cite{Binder:2001ha,Pelissetto:2000ek,Ferrenberg:2018zst}.
From Eq.~\eqref{eq:FSS_Firr}, the $n$th-order cumulant of the magnetization $M$ is expressed as
\begin{align}
    \langle M^n(t,h,L)\rangle_{\mathrm{c}}
    &= -T^{n-1} \frac{\partial^n F(t,h,L)}{\partial h^n}\Big|_T 
    \notag \\
    &= -L^{n y_h}\, \tilde{F}^{(n)}(L^{y_t}t,\, L^{y_h}h,\, \{L^{y_i}\lambda_i\}),
\end{align}
where $\tilde{F}^{(n)}$ denotes the $n$th derivative of $\tilde{F}(\tilde t,\tilde h,\{ \tilde \lambda_i\})$ with respect to $\tilde h$. At $h=0$, using the expansion 
\begin{align}
    \tilde F^{(n)}(\tilde t,0,\{\tilde \lambda_i\}) 
    = \tilde F^{(n)}(0,0,\{0\}) + (\partial_{\tilde t} \tilde F^{(n)})  \tilde{t} + (\partial_{\tilde{\lambda_1}} \tilde F^{(n)}) \tilde{\lambda_1} + (\partial_{\tilde{\lambda_2}}\tilde F^{(n)}) \tilde{\lambda_2} +  \cdots ,
\end{align}
the fourth-order Binder cumulant, Eq.~\eqref{eq:B4def}, is expanded as
\begin{align}
    B_4(t,L)
    = b_4 + c_{4} L^{y_t}t + \alpha_{4,1}L^{-\omega_1} + \alpha_{4,2}L^{-\omega_2} + \cdots,
    \label{eq:B4irr}
\end{align}
around $t=0$, where the coefficients $b_4$ and $c_4$ are the same as those in Eq.~\eqref{eq:Binder_nonlinear}. The values of $\lambda_i$ are determined by the details of the model. The expansions of $\tilde F^{(n)}$ with respect to $L^{-\omega_i}\lambda_i$ are justified for sufficiently large $L$ as $L^{-\omega_i}\lambda_i\to0$ for $L\to\infty$.

Equation~\eqref{eq:B4irr} shows that nonzero irrelevant operators lead to $L$-dependent contributions to Eq.~\eqref{eq:B4_expansion} unless $\lambda_i=0$. In particular, the value of $B_4(t,L)$ at $t=0$ is $L$ dependent due to the effect of the irrelevant operators, 
\begin{align}
    B_4(0,L) = b_4 \left( 1 + \tilde{\alpha}_{4,1} L^{-\omega_1} + \tilde{\alpha}_{4,2} L^{-\omega_2} + \cdots \right) ,
    \label{eq:B4irr0}
\end{align}
with $\tilde{\alpha}_{4,i} = \alpha_{4,i}/b_4$. Equation~\eqref{eq:B4irr0} has been employed for the high-precision numerical measurements of the value of $b_4$ in Ref.~\citen{Ferrenberg:2018zst}. The $L$-dependent terms in Eq.~\eqref{eq:B4irr0} are suppressed for $L\to\infty$, since $\omega_i>0$ for irrelevant operators. In the 3d-Ising universality class, the smallest exponent, having the most dominant contribution at large $L$, is known as $\omega_1 \simeq 0.83$~\cite{El-Showk:2014dwa}.
In Sec.~\ref{sec:FSSviolation}, we use Eq.~\eqref{eq:B4irr0} for an estimation of the value of $b_4$ (the red line in Fig.~\ref{fig:LYZRs_B4}). 

Next, let us extend the argument to the LYZRs. Equation~\eqref{eq:FSS_Firr} indicates that the contributions of the irrelevant parameters modify the FSS relation for the partition function, Eq.~\eqref{eq:FSS_Z}, as
\begin{align}
    Z(t,h,\{\lambda_i\},L^{-1}) 
    = \tilde{Z}(L^{y_t}t,\, L^{y_h}h,\, \{L^{y_i}\lambda_i\}),
    \label{eq:FSS_Zirr}
\end{align}
with the modified scaling function $\tilde{Z}$.
Since the LYZs, $h_\textrm{LY}^{(n)}(t,\{\lambda_i\},L^{-1})$, are given by the zeros of the partition function, they satisfy
\begin{align}
    Z(t,h_\textrm{LY}^{(n)}(t,\{\lambda_i\},L^{-1}),\{\lambda_i\},L^{-1})
     = \tilde Z(L^{y_t}t,L^{y_h}h_\textrm{LY}^{(n)}(t,\{\lambda_i\},L^{-1}),\{L^{y_i}\lambda_i\})=0.
     \label{eq:Z=0irr}
\end{align}
Denoting the zeros of $\tilde Z$ as 
$\tilde{h}_\textrm{LY}^{(n)}(\tilde{t},\{\tilde{\lambda}_i\})$, i.e.\
\begin{align}
    \tilde{Z}(\tilde{t},\tilde{h}_\textrm{LY}^{(n)}(\tilde{t},\{\tilde{\lambda}_i\}),\{\tilde{\lambda}_i\})=0,
\end{align}
from Eq.~\eqref{eq:Z=0irr} $h_{\rm LY}^{(n)}(t,\{\lambda_i\},L^{-1})$ and $\tilde h_{\rm LY}^{(n)}(\tilde{t},\{\tilde{\lambda}_i\})$ are related with each other as
\begin{align}
    {h}_\textrm{LY}^{(n)}({t},\{{\lambda}_i\}) 
    &= 
    L^{-y_h}\tilde{h}_\textrm{LY}^{(n)}(L^{y_t}{t},\{L^{y_i}{\lambda}_i\})
    \notag \\
    &= L^{-y_h} \Big( \tilde{h}_\textrm{LY}^{(n)}(L^{y_t}{t}) + W_{n1}^{(n)} L^{y_1}\lambda_1 + W_{n2}^{(n)} L^{y_2}\lambda_2 + \cdots \Big),
\end{align}
where on the second line we expanded $\tilde{h}_\textrm{LY}^{(n)}(L^{y_t}{t},\{L^{y_i}{\lambda}_i\})$ by $L^{y_i}\lambda_i$, and $\tilde{h}_\textrm{LY}^{(n)}(L^{y_t}{t})$ are the LYZs at $\lambda_i=0$ for all $i$, i.e. those introduced in the main text. Then, by adopting the expansion in Eq.~\eqref{eq:expansion} to $\tilde{h}_\textrm{LY}^{(n)}(L^{y_t}{t})$, we find 
\begin{align}
    R_{nm}(t,L) &= \frac{h_\textrm{LY}^{(n)}(t,\{\lambda_i\},L)}{h_\textrm{LY}^{(m)}(t,\{\lambda_i\},L)}
    = \frac{\tilde{h}_\textrm{LY}^{(n)}(L^{y_t}{t},\{L^{y_i}{\lambda}_i\})}{\tilde{h}_\textrm{LY}^{(m)}(L^{y_t}{t},\{L^{y_i}{\lambda}_i\})}
    \notag \\
    &= r_{nm} + c_{nm}L^{y_t}t + \alpha_{nm,1}L^{-\omega_1} + \alpha_{nm,2}L^{-\omega_2} + \cdots,
    \label{eq:LYZRirr}
\end{align}
where $r_{nm}$ and $c_{nm}$ are the same as those in Eq.~\eqref{eq:LYZR_nonlinear}. The coefficients from irrelevant terms $\{\alpha_{nm,i}\}$ are given by derivatives of $\tilde{h}_\textrm{LY}^{(n)}(L^{y_t}{t},\{L^{y_i}{\lambda}_i\})$ and depend on the microscopic details of the model.
We note that the effect of the irrelevant operators in Eq.~\eqref{eq:LYZRirr} is analogous to that in Eq.~\eqref{eq:B4irr} for the Binder cumulant. Substituting $t=0$ into Eq.~\eqref{eq:LYZRirr}, one finds that
\begin{align}
    R_{nm}(0,L) = r_{nm} \big(1+\tilde{\alpha}_{nm,1}L^{-\omega_1}+\tilde{\alpha}_{nm,2}L^{-\omega_2}+\cdots\big),
    \label{eqapp:rnm}
\end{align}
is satisfied at $T=T_{\rm c}$, with $\tilde{\alpha}_{nm,i}=\alpha_{nm,i}/r_{nm}$.
In Sec.~\ref{sec:FSSviolation}, we study the $L$ dependence of $R_{nm}(0,L)$ by taking the first term of Eqs.~\eqref{eqapp:rnm} having the dominant contribution for $L\to\infty$.

We notice that the critical temperature $T=T_{\rm c}$, corresponding to $t=0$, is unique in Eq.~\eqref{eq:FSS_Firr}. On the other hand, if one wants to estimate the critical temperature from the numerical simulation with a fixed $L$, one may, for example, use the peak position of the second-order cumulant. The value of $T_{\rm c}$ introduced in this way is $L$ dependent, which converges to the true critical temperature $T_{\rm c}$ in the limit $L\to\infty$. In the present study, we do not use such a definition of $T_{\rm c}$. Instead, we use $\Tc$ determined in Ref.~\citen{Ferrenberg:2018zst} or by the intersection analysis in Sec.~\ref{sec:FSSviolation}.

\begin{figure}
    \centering\includegraphics[width=0.98\linewidth]{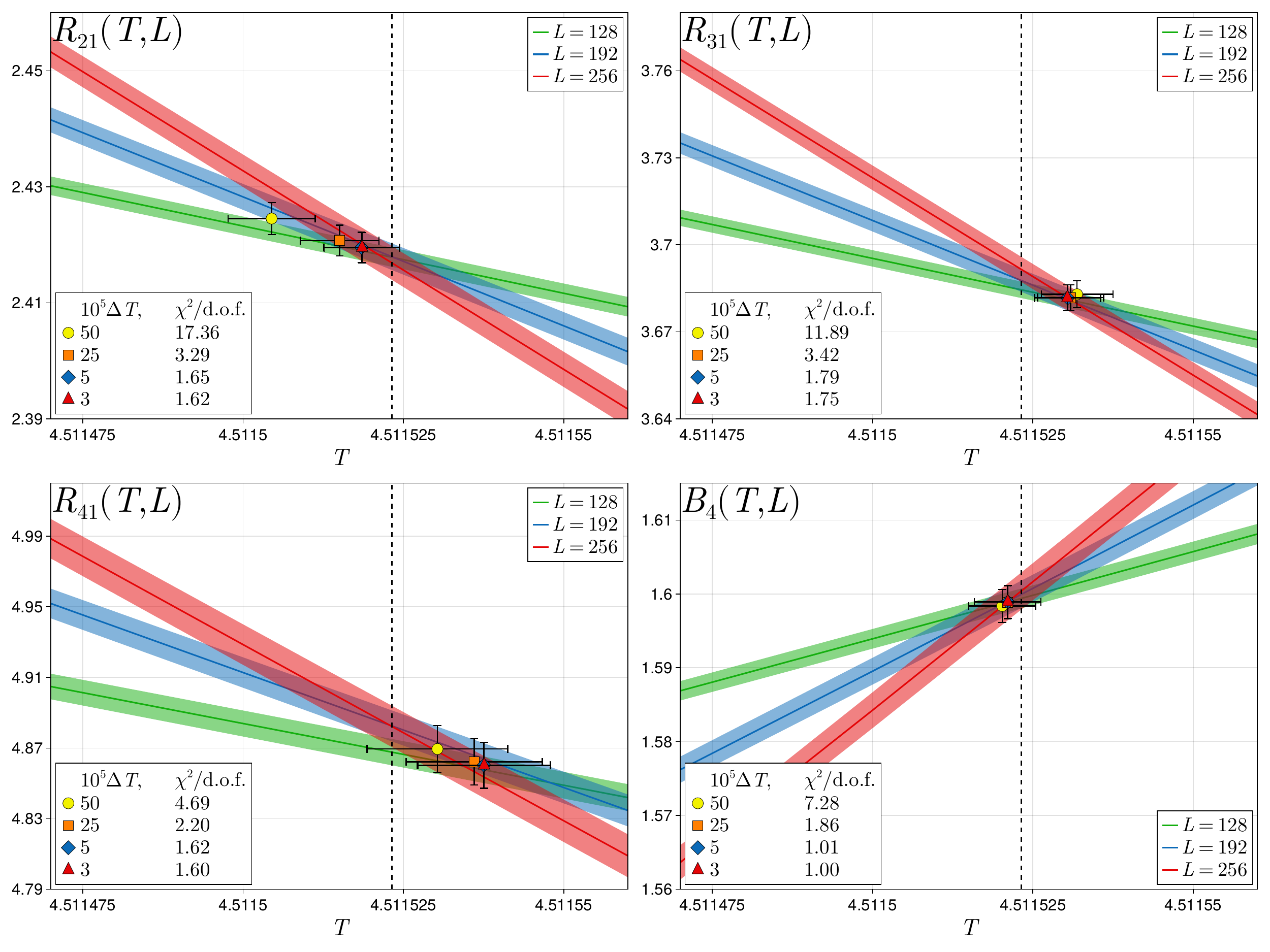}
    \caption{LYZRs $R_{n1}$ for $n=2,3,4$ and the fourth Binder cumulant $B_4$ as functions of temperature $T$ at various system sizes $L$. Colored symbols represent the fitting results with $L_\textrm{min}=128$ for various $\Delta T$.}
    \label{fig:LYZRs_T_fit}
\end{figure}

\begin{figure}
    \centering
    \includegraphics[width=0.98\linewidth]{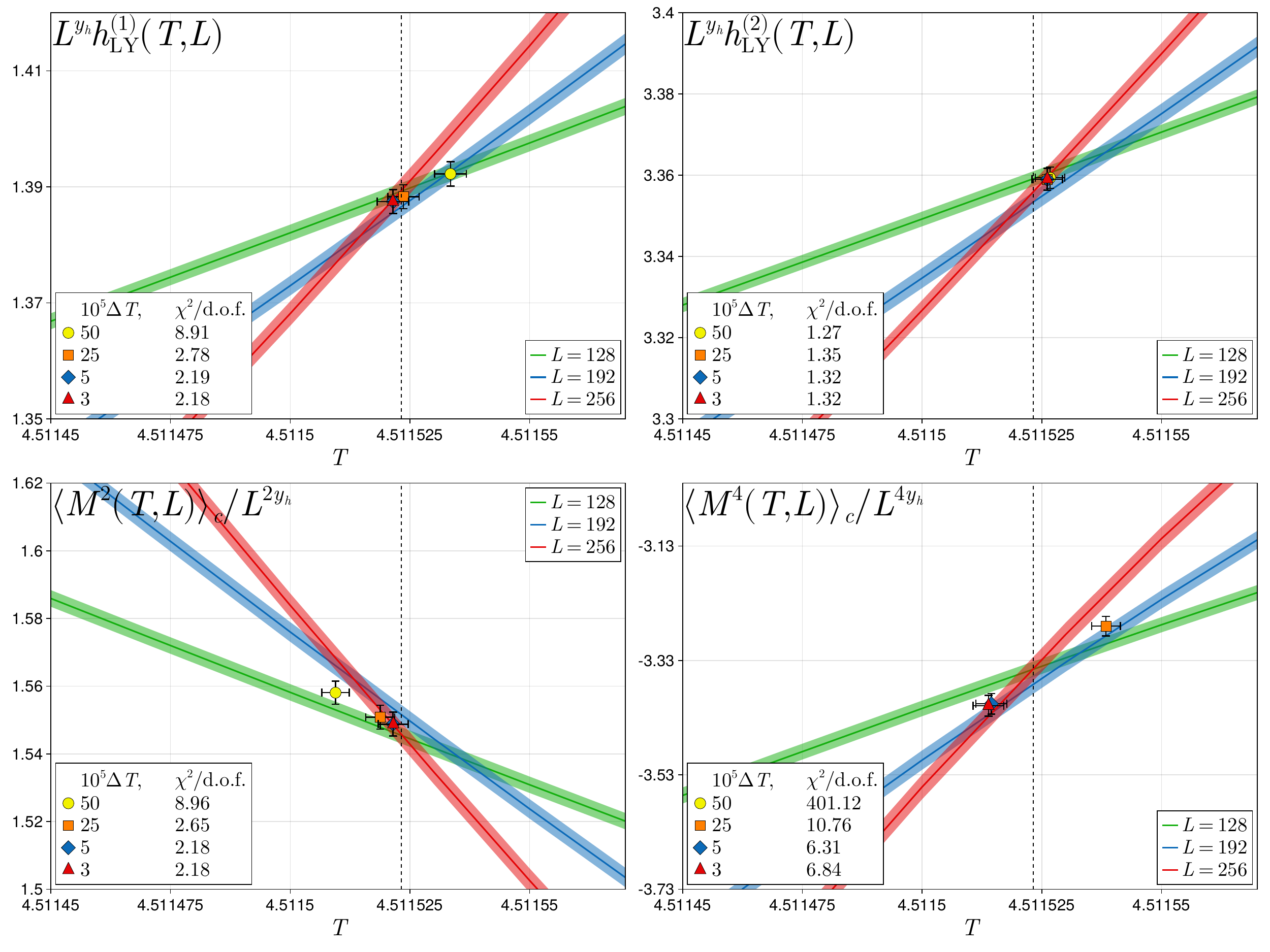}
    \caption{Rescaled LYZs $L^{y_h} h^{(n)}_\textrm{LY}$ for $n=1,2$ and rescaled cumulants $M^n/L^{ny_h}$ for $n=2,4$ for various $L$. The colored symbols are the results of fits with $L_\textrm{min}=128$ for various $\Delta T$.}
    \label{fig:LYZs_cumulant_T_fit}
\end{figure}

\section{$\Delta T$ dependence of the intersection analyses} 
\label{app:DeltaT}

In this appendix, we examine how the fitting results in Figs.~\ref{fig:LYZRs_L_fit} and~\ref{fig:LYZs_cumulant_L_fit} depend on $\Delta T$.

In Fig.~\ref{fig:LYZRs_T_fit}, we plot the LYZRs and the Binder cumulant as functions of $T$ for $L=128, 192, 256$, together with the fit results of the intersection point with fixed $L_{\rm min}=128$ but for various $\Delta T$. The procedure of the fits is the same as that explained in Sec.~\ref{sec:intersection}. From the figure, one finds that $\Delta T$ dependence is well suppressed for $\Delta T<25\times10^{-5}$, while the result for $\Delta T=50\times10^{-5}$ is clearly deviated from the other results. The results for $\Delta T\times10^5=3,5$ are almost overlapped and not distinguishable in the figure. Similar results are obtained also for other fit parameters such as $y_t$ and other choices of $L_{\rm min}$. In the main text, we thus employ $\Delta T=5\times10^{-5}$.

Next, we plot the results of the same analyses for the single-LYZ and cumulant methods in Fig.~\ref{fig:LYZs_cumulant_T_fit}. We find from the figure that the $\Delta T$ dependence of the fit results is nearly the same as in the previous analysis.

\section{Invariance of normalized curvatures}
\label{app:second}
In this appendix, we discuss the invariance of the ratios of the normalized curvatures Eq.~\eqref{eq:C_f} under reparametrizations of thermodynamic variables. We introduce new thermodynamic variables $\tau$ and $\eta$, which are analytically related to the reduced temperature $t$ and the magnetic field $h$ near the CP $(t,h)=(0,0)$, respectively, such that they can be written as 
\begin{align}
    t &= t(\tau) = t_{(1)} \tau + \frac12 t_{(2)} \tau^2 + \cdots ,
    \label{eq:t(tau)}
    \\
    h &= h(\eta) = h_{(1)} \eta + \frac12 h_{(2)} \eta^2 + \cdots ,
    \label{eq:h(eta)}
\end{align}
where we assume that $(\tau,\eta)=(0,0)$ corresponds to the CP and the linear terms are non-vanishing, $t_{(1)}\ne0$ and $h_{(1)}\ne0$.
In the following, we show that the ratios of $C_f$ are invariant under the variable transformation $(t,h)\to(\tau,\eta)$ in the $L\to\infty$ limit. Throughout this appendix, we assume the validity of the FSS, which is justified in the vicinity of the CP in the $L\to\infty$ limit.

To prove the statement, we begin with the transformation of the cumulants $\langle M^n(t,0,L)\rangle_{\rm c}=\partial^n F/\partial h^n$. In terms of the new variables, the cumulants may be defined as $\partial^n {\cal F}(\tau,0,L)/\partial \eta^n$ with the free energy as a function of the new variables ${\cal F}(\tau,\eta,L)=F(t(\tau),h(\eta),L)$. They are calculated to be
\begin{align}
    \frac{\partial^n {\cal F}(\tau,0,L)}{\partial \eta^n}
    &= h_{(1)}^n F^{(0,n)}(t(\tau),0,L) + T_n h_{(1)}^{n-2} h_{(2)} F^{(0,n-1)}(t(\tau),0,L) + \cdots
    \notag \\
    &= L^{ny_h} h_{(1)}^n \tilde{F}^{(0,n)}(L^{y_t} t(\tau),0) + T_n L^{(n-1)y_h} h_{(1)}^{n-2} h_{(2)} \tilde{F}^{(0,n-1)}(L^{y_t} t(\tau),0) + \cdots
    \notag \\
    &\xrightarrow[L\to\infty]{} L^{ny_h} h_{(1)}^n \tilde{F}^{(0,n)}(L^{y_t} t(\tau),0) ,
    \label{eq:dF/deta}
\end{align}
with $T_n=n(n-1)/2$, $F^{(i,j)}(t,h,L)=\partial^{i+j}F(t,h,L)/\partial t^i \partial h^j$, and $\tilde F^{(i,j)}(\tilde t,\tilde h)=\partial^{i+j}\tilde F(t,h,L)/\partial \tilde t^i \partial \tilde h^j$. 
The term on the far-right hand side $\tilde{F}^{(0,n)}(L^{y_t} t(\tau),0)$ is Taylor expanded as
\begin{align}
    \notag \\
    \tilde{F}^{(0,n)}(L^{y_t} t(\tau),0) 
    &= \tilde{F}^{(0,n)} + L^{y_t} t_{(1)} \tilde{F}^{(1,n)} \tau + \Big\{ L^{2y_t} t_{(1)}^2 \tilde{F}^{(2,n)} + L^{y_t} t_{(1)}t_{(2)} \tilde{F}^{(1,n)} \Big\} \tau^2 + \cdots ,
    \notag \\
    &\xrightarrow[L\to\infty]{} 
    \tilde{F}^{(0,n)} + L^{y_t} t_{(1)} \tilde{F}^{(1,n)} \tau + L^{2y_t} t_{(1)}^2 \tilde{F}^{(2,n)} \tau^2 + \cdots ,
\end{align}
where we define $\tilde{F}^{(i,j)}$ without arguments to mean $\tilde{F}^{(i,j)}(0,0)$. Keeping only the leading contribution in each order of $\tau$, Eq.~\eqref{eq:dF/deta} becomes
\begin{align}
    \frac{\partial^n {\cal F}(\tau,0,L)}{\partial \eta^n} 
    \xrightarrow[L\to\infty]{} 
    L^{ny_h} h_{(1)}^n \Big[ \tilde{F}^{(0,n)} + L^{y_t} t_{(1)} \tilde{F}^{(1,n)} \tau + L^{2y_t} t_{(1)}^2 \tilde{F}^{(2,n)} \tau^2 + \cdots \Big] .
\end{align}
The fourth-order Binder cumulant in terms of the new variables is then calculated to be
\begin{align}
    {\cal B}_4(\tau,L) 
    &\equiv \frac{\partial^4 {\cal F}(\tau,0,L)/\partial \eta^4}{(\partial^2 {\cal F}(\tau,0,L)/\partial \eta^2)^2} + 3
    \notag \\
    &\xrightarrow[L\to\infty]{} 
    b_4 + c_4L^{y_t}t^{(1)}\tau + d_4(L^{y_t}t^{(1)}\tau)^2 + \cdots,
    \label{eq:B4tau}
\end{align}
where $b_4$, $c_4$, $d_4$ are the same as Eq.~\eqref{eq:B4_expansion}. From Eq.~\eqref{eq:B4tau}, one finds
\begin{align}
    C_{{\cal B}_4} = t^{(1)}\frac{d_4}{c_4} = t^{(1)} C_{B_4}.
    \label{eq:CB4}
\end{align}
It is not difficult to obtain the same result for the other cumulant ratios.

Next, we focus on the LYZs. We consider zeros of ${\cal Z}(\tau,\eta,L)=Z(t(\tau),h(\eta),L)$ on the complex-$\eta$ plane for $\tau\in\mathbb{R}$, and denote them as $\eta=\eta_{\rm LY}^{(n)}(\tau,L)$. Since the LYZs in terms of the original variables satisfy Eq.~\eqref{eq:tildeh_LY}, $\eta_{\rm LY}^{(n)}(\tau,L)$ satisfy
\begin{align}
    L^{y_h} h\big(\eta_{\rm LY}^{(n)}(\tau,L) \big) 
    &= \tilde h_{\rm LY}^{(n)}\big( L^{y_t} t(\tau) \big)
    \notag \\
    &= i\Big[ X_n + Y_n L^{y_t} t(\tau) + Z_n (L^{y_t} t(\tau))^2 + \cdots \Big]
    \notag \\
    &= i\Big[ X_n + L^{y_t} Y_n t_{(1)} \tau + \Big\{ L^{2y_t} Z_n t_{(1)}^2 + L^{y_t} Y_n \frac{t_{(2)}}2 \Big\} \tau^2 \cdots \Big] ,
    \label{eq:h(eta)XYZ}
\end{align}
Denoting the Taylor expansion of $\eta_{\rm LY}^{(n)}(\tau,L)$ as 
\begin{align}
    \eta_{\rm LY}^{(n)}(\tau,L)
    = {\cal X}_n + {\cal Y}_n \tau + {\cal Z}_n \tau^2 + \cdots ,
    \label{eq:eta_exp}
\end{align}
$h(\eta_{\rm LY}^{(n)}(\tau,L) )$ on the left-hand side is calculated to be
\begin{align}
    h(\eta_{\rm LY}^{(n)}(\tau,L) ) 
    &= h_{(1)} \eta_{\rm LY}^{(n)}(\tau,L) + \frac{h_{(2)}}2 (\eta_{\rm LY}^{(n)}(\tau,L) )^2 + \cdots 
    \notag \\
    &= h_{(1)} {\cal X}_n + \frac{h_{(2)}}2 {\cal X}_n^2 + \Big( h_{(1)} {\cal Y}_n + h_{(2)} {\cal X}_n {\cal Y}_n \Big) \tau 
    \notag \\
    &\phantom= + \Big( h_{(1)} {\cal Z}_n + h_{(2)} {\cal X}_n {\cal Z}_n + \frac{h_{(2)}}2 {\cal Y}_n^2 \Big) \tau^2 + \cdots .
    \label{eq:h(eta)exp}
\end{align}
Substituting Eq.~\eqref{eq:h(eta)exp} into the left-hand side of Eq.~\eqref{eq:h(eta)XYZ} and comparing terms in each order of $\tau$, we find
\begin{align}
    h_{(1)} {\cal X}_n + \frac{h_{(2)}}2 {\cal X}_n^2 
    &= L^{-y_h} iX_n ,
    \label{eq:eta0}
    \\
    h_{(1)} {\cal Y}_n + h_{(2)} {\cal X}_n {\cal Y}_n 
    &= L^{y_t-y_h} iY_n t_{(1)} ,
    \label{eq:eta1}
    \\
    h_{(1)} {\cal Z}_n + h_{(2)} {\cal X}_n {\cal Z}_n + \frac{h_{(2)}}2 {\cal Y}_n^2 &= L^{2y_t-y_h} iZ_n t_{(1)}^2 + L^{y_t-y_h} iY_n \frac{t_{(2)}}2 .
    \label{eq:eta2}
\end{align}

The coefficients ${\cal X}_n$, ${\cal Y}_n$, ${\cal Z}_n$ in Eq.~\eqref{eq:eta_exp} must satisfy these equations. 
Equation~\eqref{eq:eta0} and the fact that ${\cal X}_n\to0$ for $L\to\infty$ lead to 
\begin{align}
    L^{y_h} h_{(1)} {\cal X}_n \xrightarrow[L\to\infty]{} iX_n .
    \label{eq:Xn}
\end{align}
Then, by taking the leading terms for $L\to\infty$ and using Eq.~\eqref{eq:Xn}, one finds
\begin{align}
    L^{y_h} h_{(1)} {\cal Y}_n &\xrightarrow[L\to\infty]{} L^{y_t} i Y_n t_{(1)} ,
    \\
    L^{y_h} h_{(1)} {\cal Z}_n &\xrightarrow[L\to\infty]{} 
    L^{2y_t} i Z_n t_{(1)}^2 ,
    \label{eq:Zn}
\end{align}
where we used $y_t-y_h<0$ to obtain the last equality.

From Eqs.~\eqref{eq:Xn}--\eqref{eq:Zn}, the LYZRs in the new coordinates are calculated to be
\begin{align}
    {\cal R}_{nm}(\tau,L) = \frac{\eta_{\rm LY}^{(n)}(\tau,L)}{\eta_{\rm LY}^{(m)}(\tau,L)}
    \xrightarrow[L\to\infty]{} 
    r_{nm} + L^{y_t} c_{nm} t_{(1)} \tau + d_{nm} (L^{y_t} t_{(1)} \tau)^2 + \cdots,
\end{align}
where $r_{nm}$, $c_{nm}$, $d_{nm}$ are the same as those in Eq.~\eqref{eq:LYZRs_expansion}, which means that
\begin{align}
    C_{{\cal R}_{nm}} = t_{(1)} \frac{d_{nm}}{c_{nm}} = t_{(1)} C_{R_{nm}} .
    \label{eq:CRnm}
\end{align}
Comparing Eqs.~\eqref{eq:CB4} and~\eqref{eq:CRnm}, we find 
\begin{align}
    \frac{C_{{\cal R}_{nm}} }{C_{{\cal B}_4}} = \frac{C_{R_{nm}} }{C_{B_4}} ,
\end{align}
that is, the ratio is invariant under the variable transformation~\eqref{eq:t(tau)} and~\eqref{eq:h(eta)}. 

Finally, we notice that the ratios of the normalized curvatures are not invariant under more general variable transformations that allow the mixing of two variables, i.e.
\begin{align}
    \begin{pmatrix}
        t \\ h
    \end{pmatrix}
    =
    \begin{pmatrix}
        a_{11} & a_{12} \\ a_{21} & a_{22}
    \end{pmatrix}
    \begin{pmatrix}
        \tau \\ \eta
    \end{pmatrix} 
    + 
    \begin{pmatrix}
        b_{11} & b_{12} & b_{13} \\
        b_{21} & b_{22} & b_{23}
    \end{pmatrix} 
    \begin{pmatrix}
        \tau^2 \\ \tau\eta \\ \eta^2
    \end{pmatrix} ,
\end{align}
with nonzero mixing elements such as $a_{12}$ and $a_{21}$.

\bibliographystyle{jpsj}
\bibliography{bibliography}

\end{document}